\documentclass{emulateapj}

\shorttitle{DISSECTING THE GRAVITATIONAL LENS B1608+656. I.}
\shortauthors{Suyu et al.}

\usepackage{natbib}
\usepackage{latexsym}
\usepackage{amssymb}

\newcommand{\bd}{\begin{displaymath}}
\newcommand{\ed}{\end{displaymath}}
\newcommand{\be}{\begin{equation}}
\newcommand{\ee}{\end{equation}}
\newcommand{\beaa}{\begin{eqnarray*}}
\newcommand{\eeaa}{\end{eqnarray*}}
\newcommand{\bea}{\begin{eqnarray}}
\newcommand{\eea}{\end{eqnarray}}

\def\HST{\textit{HST}{}}

\newcommand{\boldsymbol}[1]{\mbox{\boldmath{${#1}$}}}

\newcommand{\bmath}{\vec}

\def\dataVec{\boldsymbol{d}}
\def\data{d}
\def\responseSet{\boldsymbol{\mathsf{f}}}
\def\response{\mathsf{f}}
\def\regSet{\boldsymbol{\mathsf{g}}}

\def\srVec{\boldsymbol{s}}
\def\sr{s}

\def\srMPVec{\boldsymbol{s}_{\mathrm{MP}}}

\def\noiseVec{\boldsymbol{n}}
\def\noise{n}

\def\imCM{\boldsymbol{\mathsf{C}}_{\mathrm{D}}}

\def\hessD{\boldsymbol{\mathsf{B}}}
\def\hessS{\boldsymbol{\mathsf{C}}}
\def\hessM{\boldsymbol{\mathsf{A}}}

\def\dI{\delta I}
\def\dIVec{\boldsymbol{\delta I}}
\def\dpsi{\delta\psi}
\def\dpsiVec{\boldsymbol{\delta \psi}}
\def\dpsiMPVec{\boldsymbol{\delta \psi}_{\mathrm{MP}}}
\def\PRmatSet{\boldsymbol{\mathsf{t}}}

\def\regSetdpsi{\boldsymbol{\mathsf{g}_{\dpsi}}}

\def\blurSet{\boldsymbol{\mathsf{B}}}

\def\dustSet{\boldsymbol{\mathsf{K}}}
\def\dust{\mathsf{K}}

\def\lensSet{\boldsymbol{\mathsf{L}}}

\def\glightVec{\boldsymbol{l}}

\begin{document}

\title{DISSECTING THE GRAVITATIONAL LENS B1608+656. I. LENS POTENTIAL RECONSTRUCTION\altaffilmark{*}}


\author{S.~H.~Suyu\altaffilmark{1,2,3},     
        P.~J.~Marshall\altaffilmark{4},
        R.~D.~Blandford\altaffilmark{1,2}, 
        C.~D.~Fassnacht\altaffilmark{5},
        L.~V.~E.~Koopmans\altaffilmark{6}, \\
        J.~P.~McKean\altaffilmark{5,7}, and
        T.~Treu\altaffilmark{4,8}}

\email{suyu@astro.uni-bonn.de}
\altaffiltext{*}{Based in part on observations made with the NASA/ESA \textit{Hubble Space Telescope}, obtained at the Space Telescope Science Institute, which is operated by the Association of Universities for Research in Astronomy, Inc., under 
NASA contract NAS 5-26555. These observations are associated with program 
GO-10158.}
\altaffiltext{1}{Theoretical Astrophysics, 103-33, California Institute of Technology, Pasadena, CA, 91125, USA}
\altaffiltext{2}{Kavli Institute for Particle Astrophysics and Cosmology, Stanford University, PO Box 20450, MS 29, Stanford, CA 94309, USA}
\altaffiltext{3}{Argelander-Institut f\"{u}r Astronomie, Auf dem H\"{u}gel 71, D-53121 Bonn, Germany}
\altaffiltext{4}{Department of Physics, University of California, Santa Barbara, CA 93106-9530, USA}
\altaffiltext{5}{Department of Physics, University of California at Davis, 1 Shields Avenue, Davis, CA 95616, USA}
\altaffiltext{6}{Kapteyn Astronomical Institute, University of Groningen, P.O.Box800, 9700AV Groningen, The Netherlands}
\altaffiltext{7}{Max-Planck-Institut f\"{u}r Radioastronomie, Auf dem H\"{u}gel 69, D-53121 Bonn, Germany}
\altaffiltext{8}{Sloan Fellow, Packard Fellow}

\begin{abstract}
  Strong gravitational lensing is a powerful technique for probing
  galaxy mass distributions and for measuring cosmological parameters.
  Lens systems with extended source-intensity distributions are
  particularly useful for this purpose since they provide additional
  constraints on the lens potential (mass distribution).  We present a
  pixelated approach to modeling the lens potential and
  source-intensity distribution simultaneously.  The method makes iterative and
  perturbative corrections to an initial potential model.  For systems
  with sources of sufficient extent such that the separate lensed
  images are connected by intensity measurements, the accuracy in the
  reconstructed potential is solely limited by the quality of the
  data.  We apply this potential reconstruction technique to deep 
  \textit{Hubble Space Telescope}
  observations of B1608+656, a four-image gravitational lens system
  formed by a pair of interacting lens galaxies.  We present a
  comprehensive Bayesian analysis of the system that takes into
  account the extended source-intensity distribution, dust extinction,
  and the interacting lens galaxies.  Our approach allows us to
  compare various models of the components of the lens system, which
  include the point-spread function (PSF), dust, lens galaxy light,
  source-intensity distribution, and lens potential.  Using optimal
  combinations of the PSF, dust, and lens galaxy light models, we
  successfully reconstruct both the lens potential and the extended
  source-intensity distribution of B1608+656.  The resulting
  reconstruction can be used as the basis of a measurement of the
  Hubble constant.  As an illustration of the astrophysical
  applications of our method, we use our reconstruction of the
  gravitational potential to study the relative distribution of mass
  and light in the lensing galaxies. We find that the mass-to-light
  ratio for the primary lens galaxy is $(2.0\pm0.2)h \rm{\, M_{\sun}
    \, L_{B,\sun}^{-1}}$ within the Einstein radius ($3.9 h^{-1}\,
  \rm{kpc}$), in agreement with what is found for noninteracting lens
  galaxies at the same scales.
\end{abstract}



\section{Introduction}
\label{sec:intro}

\setcounter{footnote}{8}

Strong gravitational lens systems provide a tool for probing galaxy
mass distributions (independent of their light profiles) and for
measuring cosmological parameters \citep*[e.g.][ and references
therein]{KochanekEtal06}.  Lens systems with extended source-intensity
distributions are of special interest because they provide additional
constraints on the lens potential (and hence the surface mass density)
due to surface brightness conservation.  In this case, simultaneous
determination of the source-intensity distribution and the lens
potential is needed.  To describe either the source-intensity or the
lens potential/mass distribution, there are two approaches in the
literature: (1) ``parametric,'' or better, ``simply parameterized,''
using simple, physically motivated functional forms described by a few
($\sim 10$) parameters (e.g.,
\citeauthor{Kochanek91} \citeyear{Kochanek91};
\citeauthor{KneibEtal96} \citeyear{KneibEtal96}; 
\citeauthor{Keeton01} \citeyear{Keeton01}; 
\citeauthor{Marshall06} \citeyear{Marshall06};
\citeauthor{JulloEtal07} \citeyear{JulloEtal07}),
and (2) pixel-based (``pixelated,'' or ``free-form,'' or sometimes, 
inaccurately, ``nonparametric'') modeling on a grid, which has been 
done for both the source intensity (e.g.,
\citeauthor{WallingtonEtal96} \citeyear{WallingtonEtal96};
\citeauthor{WarrenDye03} \citeyear{WarrenDye03};
\citeauthor{TreuKoopmans04} \citeyear{TreuKoopmans04};
\citeauthor{DyeWarren05} \citeyear{DyeWarren05};
\citeauthor{Koopmans05} \citeyear{Koopmans05};
\citeauthor{BrewerLewis06} \citeyear{BrewerLewis06};
\citeauthor{SuyuEtal06} \citeyear{SuyuEtal06};
\citeauthor{WaythWebster06} \citeyear{WaythWebster06};
\citeauthor{DyeEtal08} \citeyear{DyeEtal08}) 
and the lens potential/mass distribution (e.g., 
\citeauthor{WilliamsSaha00} \citeyear{WilliamsSaha00};
\citeauthor{BradacEtal05} \citeyear{BradacEtal05};
\citeauthor{Koopmans05} \citeyear{Koopmans05};
\citeauthor{SahaEtal06} \citeyear{SahaEtal06};
\citeauthor{SuyuBlandford06} \citeyear{SuyuBlandford06};
\citeauthor{JeeEtal07} \citeyear{JeeEtal07};
\citeauthor{VegettiKoopmans08} \citeyear{VegettiKoopmans08}). 
Most of the developed lens modeling methods are simply parameterized.
In particular, for the measurement of the Hubble constant, lens
potential/mass models prior to \citet{SahaEtal06} have been
simply parameterized because most of the strong lens systems with time
delay measurements have only point sources (as opposed to extended
sources) to constrain the lens potential/mass distribution.  A precise
measurement of the value of $H_0$ is important for testing the flat
$\Lambda$-cold dark matter (CDM) model and studying dark energy.  The cosmic microwave
background (CMB) allows determination of cosmological parameters with
high accuracy with the exception of $H_0$ \citep[e.g.][]{KomatsuEtal08}.  An
independent measurement of $H_0$ to better than a few percent
precision provides the single most useful complement to the CMB for
dark energy studies \citep{Hu05}.

\citet{Koopmans05} developed a method for pixelated source-intensity
and lens potential reconstruction that is based on the potential
correction scheme proposed by \citet*{BlandfordEtal01}.  This
pixelated potential reconstruction method is applicable to lens
systems with extended source-intensity distributions.  Pixel-based
modeling has the advantage over simply-parameterized modeling in the
flexibility in the parametrization.  This is especially important in
complex lens systems (e.g. multicomponent source galaxies or multiple
lens galaxies) where simply-parameterized models may become
inadequate.  Furthermore, pixel-based modeling has the capabilities of
detecting dark matter substructures \citep{Koopmans05, VegettiKoopmans08}.

In this paper, we present a lens modeling technique that is similar to
that of \citet{Koopmans05}, but in a Bayesian framework to allow quantitative
comparison between various source intensity and lens potential models.
The point-spread function (PSF), lens galaxy light, and dust models
are also incorporated in this scheme.  Therefore, this method provides
a way to rank these data models (with the five interdependent
components: source-intensity distribution, lens potential, PSF, lens
galaxy light and dust) quantitatively.  There are also propagation
effects due to structures along the line of sight (LOS), but we ignore this
for now and characterize this in a forthcoming paper (Paper II).

We choose to reconstruct the lens potential instead of the surface
mass density because (1) it is the quantity that directly relates to
the cosmological parameters via the time delays and angular diameter
distance ratios, and (2) the surface mass density can, in principle, be
easily obtained by differentiation.  In contrast,
\citet{WilliamsSaha00} and \citet{SahaEtal06} pixelized the surface
mass density.  Since the surface mass density over the entire lens
plane is required in the integral for obtaining the lens potential,
the conversion of the (finite) gridded mass density to the lens
potential is not straightforward.

We apply the pixelated potential reconstruction method to B1608+656
\citep{MyersEtal95}, a quadruple image gravitational lens system with
an extended source at $z_{\rm s}= 1.394$ \citep{FassnachtEtal96}, and
two interacting galaxy lenses at $z_{\rm d}= 0.6304$
\citep{MyersEtal95}.  B1608+656 is special in that it is the only
four-image gravitational lens systems with all three independent time
delays between the images measured with errors of only a few percent
\citep{FassnachtEtal99,FassnachtEtal02}.  Thus, it provides a great
opportunity to measure the Hubble constant, which is the subject of
Paper II.  To obtain the Hubble constant to high precision, an
accurate lens potential model is crucial.  \citet{KoopmansFassnacht99}
modeled this system using simply-parameterized lens potentials, but
did not account for the presence of dust and the extended source
intensity.  \citet{KoopmansEtal03} improved on the
simply-parameterized modeling of the lens potential with the treatment
of dust, the use of a simply-parameterized extended source-intensity
distribution, and the inclusion of constraints from stellar dynamics.
However, \citet{SuyuBlandford06} showed that this most up-to-date
simply-parameterized lens model in \citet{KoopmansEtal03} fails
certain tests such as the crossing of the critical curve through the
saddle point of the figure-eight-shaped intensity contour of the
merging images.  This suggests that the pixelated potential method may
be better suited than a simply-parameterized method for the two
interacting galaxies.  In this paper, we deliver a comprehensive
analysis of the B1608+656 system that incorporates the effects of the
extended source intensity, presence of dust, and interacting lenses.
The dissection of B1608+656 allows us to study the relative
distribution of mass and light in the interacting lens galaxies.
 
The outline of the paper is as follows.  In Section
\ref{sec:PPRMethod}, we introduce the pixelated potential
reconstruction method.  We demonstrate the method using simulated data
in Section \ref{sec:PPRMethod:demo} and generalize the method to real
data in Section \ref{sec:PPRMethod:realData}.  The remaining sections
of the paper target B1608+656.  In Section \ref{sec:ImProc}, we
summarize the \textit{Hubble Space Telescope} (\HST) observations of
B1608+656 and present the image processing.  In Section
\ref{sec:PotRec:B1608}, we show a pixelated potential reconstruction
of B1608+656.  Finally, in Section \ref{sec:B1608prop}, we comment on
the mass-to-light (M/L) ratio in B1608+656 based on the results of our
lensing analysis.  In Paper II, we use the resulting potential
reconstruction of B1608+656 together with a study of the lens
environment to infer the value of the Hubble constant.

Throughout this paper, we assume a flat $\Lambda$-CDM universe with
$\Omega_{m}=0.26$, $\Omega_{\Lambda}=0.74$, and $H_0 = 100h \mathrm{\,
  km\, s^{-1}\, Mpc^{-1}}$ \citep{KomatsuEtal08}.  For the lens and
source redshifts in B1608+656, $1''$ on the sky corresponds to $4.9
h^{-1}\, \rm{kpc}$ on the lens plane and $6.1 h^{-1}\, \rm{kpc}$ on
the source plane.


\section{Pixelated potential reconstruction}
\label{sec:PPRMethod}
In the following subsections, we present the pixelated potential
reconstruction method.  Section \ref{sec:PPRMethod:method} contains
the formalism of the method, and Section \ref{sec:PPRMethod:matrix} is
a practical implementation of the method.


\subsection{Formalism for iterative and perturbative potential corrections}
\label{sec:PPRMethod:method}
The iterative and perturbative potential correction scheme for lens
systems with extended sources was first suggested by
\citet{BlandfordEtal01} and studied by \citet{Koopmans05},
\citet{SuyuBlandford06}, and recently by \citet{VegettiKoopmans08}.  The pixelated potential reconstruction
method that we present here is similar to that in \citet{Koopmans05}
but differs in the numerical details and our use of Bayesian analysis,
which allows for model comparison.  The method in \citet{VegettiKoopmans08} 
is also based on Bayesian analysis and has adaptive gridding on the 
source plane.  In the rest of the section, we
briefly outline the theory of pixelated potential reconstruction.

The central concept for this method is to start with an initial lens
potential model and to correct it, perturbatively and iteratively, to
obtain an estimate of the true lens potential.  The initial lens
potential will usually be simply-parameterized (to allow faster
convergence with a smaller number of parameters) and ideally would be
close to the true potential.  It will then be refined via corrections
on a grid of pixels.  Obtaining the parameter values in the initial
lens potential is often a nonlinear process; in contrast, the
potential correction in each iteration is a linear inversion.

One way to think about this procedure is to observe that in a
perfectly observed image, nested intensity contours in the source
plane map onto multiple regions of the image plane.  Intensity is
preserved by the lens and so the map is from a set of single source
contours to the corresponding image contours.  The only freedom that
we have is to slide image points along the contours.  Using the fact
that the deflection field is curl-free effectively removes this
freedom.  What we describe is a procedure to determine this map that
takes into account a finite PSF, dust extinction, and source-intensity
contamination by the lens galaxy light.  In Paper II, we also include
the influence of propagation effects.

To keep the formalism simple for the moment, let us ignore the effects
of the PSF, dust extinction, and lens galaxy light.  Let
$\bmath{\theta}$ be the coordinates on the image plane and
$\bmath{\beta}$ be the coordinates on the source plane.  Let
$I_{\rm d}(\bmath{\theta})$ be the observed image intensity of a lensed
extended source, and let $\psi(\bmath{\theta})$ be an initial scaled
surface potential model\footnote{$\psi$ includes the distance ratio.}
for the lens system.  Given $\psi(\bmath{\theta})$, one can obtain the
best-fitting source-intensity distribution \citep[e.g.,][and
references therein]{ SuyuEtal06}.  Let
$I_{\rm s}(\bmath{\theta}(\bmath{\beta}))$ be the source intensity
translated to the image plane via the potential model,
$\psi({\bmath{\theta}})$, where $\bmath{\theta}$ and $\bmath{\beta}$
are related via the lens equation $\bmath{\theta}=\bmath{\beta}-
\bmath{\nabla}\psi({\bmath{\theta}})$.  We define the intensity
deficit (also known as the image residual) on the image plane by
\be \label{eq:dI} 
\delta I(\bmath{\theta}) = I_{\rm d}(\bmath{\theta}) - I_{\rm s}(\bmath{\theta}(\bmath{\beta})).
\ee

Suppose the initial lens potential model is perturbed from the true
potential, $\psi_{0}(\bmath{\theta})$, by $\delta \psi(\bmath{\theta})$:
\be \label{eq:potmodel}
\psi(\bmath{\theta}) = \psi_{0}(\bmath{\theta}) + \delta \psi(\bmath{\theta}).
\ee
For a given image (fixed $I_{\rm d}(\bmath{\theta})$) and the initial
potential model $\psi(\bmath{\theta}$), we can relate the intensity
deficit to the potential perturbation $\delta \psi(\bmath{\theta})$ by
\be \label{eq:pertEq}
\delta I(\bmath{\theta}) = \frac{\partial I_{\rm s}(\bmath{\beta})}{\partial \bmath{\beta}} \boldsymbol{\cdot} \frac {\partial \delta \psi(\bmath{\theta})}{\partial \bmath{\theta}},
\ee
to first order in $\delta \psi(\bmath{\theta})$  
(see e.g., \citet{SuyuBlandford06} for details). 
The source-intensity
gradient ${\partial I_{\rm s}(\bmath{\beta})}/{\partial \bmath{\beta}}$
implicitly depends on the potential model $\psi(\bmath{\theta})$ since
the source position $\bmath{\beta}$ (where the gradient is evaluated)
is related to $\psi(\bmath{\theta})$ via the lens equation.  We can solve
Equation (\ref{eq:pertEq}) for $\delta \psi(\bmath{\theta})$ given the
intensity deficit and source-intensity gradients, update the initial
(or previous iteration's) potential model, and repeat the process of
source-intensity reconstruction and potential correction until the
potential converges to the true solution with zero intensity deficit.
In Section \ref{sec:PPRMethod:matrix}, we focus on solving Equation
(\ref{eq:pertEq}).


\subsection{Implementation of pixelated potential reconstruction}
\label{sec:PPRMethod:matrix}

\subsubsection{Probability theory}
\label{sec:PPRMethod:matrix:probTheory}
The first step in solving Equation (\ref{eq:pertEq}) for the potential
perturbation is to obtain the source-intensity gradients and the
intensity deficit, which appear in the correction equation.  We follow
\citet{SuyuEtal06} to obtain the source-intensity distribution on a
grid of pixels given the current iteration's lens potential model.  In
this source reconstruction approach, the data (observed image) are
described by the vector $\data_j$, where $j=1,\ldots, N_{\rm d}$ and
$N_{\rm d}$ is the number of data pixels.  The source intensity is
described by the vector $\sr_i$, where $i=1,\ldots, N_{\rm s}$ and
$N_{\rm s}$ is the number of source-intensity pixels.  The observed
image is related to the source intensity via $\data_j =
\response_{ji}\sr_i + \noise_j$, where $\response_{ji}$ is the
so-called blurred lensing operator (mapping matrix) that incorporates
the lens potential (which governs the deflection of light rays) and
the PSF (blurring),\footnote{Dust extinction, if present, is also
  included in this mapping matrix $\response_{ji}$} and $\noise_j$ is
the noise in the data characterized by the covariance matrix $\imCM$.
In the inference of $\sr_i$, we impose a prior on $\sr_i$, which can
be thought of as ``regularizing'' the parameters $\sr_i$ to avoid
overfitting to the noise in the data.  Following \citet{SuyuEtal06},
we use quadratic forms of the regularization (specifically,
zeroth-order, gradient, and curvature forms of regularization).
The Bayesian inference of the source-intensity distribution ($\sr_i$)
given the observed image ($\data_j$) is a linear inversion and is a
solved problem.  Having obtained the source intensity, we can
calculate the intensity deficit and source-intensity gradients.
 
We pixelize the lens potential to allow for a flexible parametrization
scheme.  To solve Equation (\ref{eq:pertEq}), we cast it into a matrix
equation and invert the linear system.  To write Equation
(\ref{eq:pertEq}) in a matrix form, we discretize the lens potential
on a rectangular grid of $N_{\rm p}$ pixels (which is 
less than the number of data pixels $N_{\rm d}$ so that the
potential and source-intensity pixels are not underconstrained) and
denote the potential perturbation by $\dpsi_i$ where $i=1,\ldots, N_{\rm
  p}$.  The intensity deficit on the image grid is $\dI_j=\data_j -
\response_{ji}\sr_i$ where $j=1,\ldots, N_{\rm d}$ (using the notation
from source-intensity reconstruction, $\dataVec$, $\responseSet$ and
$\srVec$ are the data vector, the blurred lensing operator, and the
source-intensity vector, respectively).  Equation (\ref{eq:pertEq})
now becomes 
\be
\label{eq:pertEqMat}
\dIVec = \PRmatSet \dpsiVec + \noiseVec,
\ee
where $\PRmatSet$ is a $N_{\rm d} \times N_{\rm p}$ matrix which
incorporates the PSF, the source-intensity gradient, and the gradient
operator that acts on $\dpsiVec$ (see the appendix for the explicit
form of $\PRmatSet$), and $\noiseVec$ is the noise in the data.  The
above equation is equivalent to
\be
\label{eq:djWithPert}
\dataVec = \responseSet \srVec + \PRmatSet \dpsiVec + \noiseVec.
\ee

We can infer the potential corrections $\dpsiVec$ given the data
$\dataVec$, source intensity $\srVec$, and source-intensity gradients
that are encoded in $\PRmatSet$.  In the inference, we impose a prior
on $\dpsiVec$.
The posterior probability distribution is
\be
\label{eq:dpsiPosterior}
\overbrace{P(\dpsiVec|\dataVec, \responseSet, \srVec, \PRmatSet, \mu, \regSet_{\dpsi})}^{\rm{posterior}} = \frac{ \overbrace{P(\dataVec|\dpsiVec, \PRmatSet, \responseSet, \srVec)}^{\rm{likelihood}} \overbrace{P(\dpsiVec|\mu, \regSetdpsi)}^{\rm{prior}}}{\underbrace{P(\dataVec|\responseSet, \srVec, \PRmatSet, \mu, \regSetdpsi)}_{\rm{evidence}}}, 
\ee
where $\mu$ and $\regSetdpsi$ are the (fixed) strength and form of
regularization for the potential correction inversion,
and all irrelevant (in)dependences have been dropped.  
Modeling the noise as
Gaussian, the likelihood is
\be
\label{eq:dpsiLikelihood}
P(\dataVec|\dpsiVec, \PRmatSet, \responseSet, \srVec) = \frac{\exp(-E_{\rm{D}}(\dataVec|\dpsiVec, \PRmatSet, \responseSet, \srVec))}{Z_{\rm{D}}}, 
\ee
where 
\bea
E_{\rm{D}}(\dataVec|\dpsiVec, \PRmatSet, \responseSet, \srVec) &=& \frac{1}{2}(\dataVec-\responseSet\srVec - \PRmatSet\dpsiVec)^{\rm{T}} \imCM^{-1} \nonumber \\
& & {\ \ \ \ \ \ \ } (\dataVec-\responseSet\srVec - \PRmatSet\dpsiVec)\\
& = & \frac{1}{2}\chi^2
\eea
and $Z_{\rm{D}}$ is the normalization for the probability.  We express
the prior in the following form:
\be
P(\dpsiVec|\mu, \regSetdpsi) = \frac{\exp(-\mu E_{\mathrm{\dpsi}}(\dpsiVec|\regSetdpsi))}{Z_{\dpsi}(\mu)}.
\ee
We use quadratic forms of the regularizing function $E_{\mathrm{\dpsi}}$.  In
particular, we use the curvature form of regularization (see, for example,
Appendix A of \citet{SuyuEtal06} for an explicit expression of the curvature
form of regularization).  We use this regularization instead of the zeroth-order
or gradient forms because the lens potential should in general be smooth, being
the \textit{integral} of the surface mass density.  Curvature regularization in
the potential corrections effectively corresponds to zeroth-order
regularization in the surface mass density corrections.  This implies
a prior preference toward zero surface mass density corrections, thus 
suppressing the addition of mass to the initial mass model unless the data
require it.

Maximizing the posterior of parameters $\dpsiVec$, we obtain the most probable
solution
\be 
\label{eq:dpsiMP}
\dpsiMPVec = \hessM^{-1} \boldsymbol{D},
\ee
where
\bea
\label{eq:hessDefs}
\nonumber \hessM & = & \hessD + \mu\hessS, \\
\nonumber \hessD & \equiv & \nabla \nabla E_{\mathrm{D}}(\dpsiVec)  = \PRmatSet^{\mathrm{T}} \imCM^{-1} \PRmatSet, \\
\nonumber \hessS & \equiv & \nabla \nabla E_{\mathrm{\dpsi}}(\dpsiVec), \\ 
\nonumber \boldsymbol{D} & = & \PRmatSet^{\mathrm{T}} \imCM^{-1} (\dataVec-\responseSet\srVec), \\
\nonumber \rm {and\ }  \nabla & \equiv & \frac{\partial}{\partial \dpsiVec}.
\eea
The matrices $\hessM$, $\hessD$ and $\hessS$ have dimensions $N_{\rm p}
\times N_{\rm p}$ and are, by definition, the Hessians of the exponential
arguments in the posterior, the likelihood, and the prior probability
distributions, respectively.

As discussed in detail in, for example, \citet{MacKay92} and
\citet{SuyuEtal06}, the evidence is irrelevant in the first level of
inference where we maximize the posterior of parameters $\dpsiVec$ to
obtain the most probable parameters $\dpsiMPVec$.  However, the
evidence is crucial for the second level of inference for model
comparison, where a model incorporates the lens potential, PSF, and
regularizations of both the source intensity and the potential
correction.  If we assert that models are equally probable a
priori, then the evidence gives the relative probability of the
model given the data.  In other words, the ratio in the evidence
values of two models tells us how much more probable the first model
is relative to the second model, if we assume that the two models are
a priori equally probable.  
Since the evidence gives only the relative probability, the data set
needs to be kept the same for model comparison.

The posterior ($P(\dpsiVec|\dataVec, \responseSet, \srVec, \PRmatSet,
\mu, \regSet_{\dpsi})$) and the evidence ($P(\dataVec|\responseSet,
\srVec, \PRmatSet, \mu, \regSetdpsi)$) in Equation
(\ref{eq:dpsiPosterior}) are conditional on the source-intensity
distribution.  Ideally, we would have an expression of the posterior
for both $\srVec$ and $\dpsiVec$: $P(\srVec, \dpsiVec|\dataVec,
\responseSet, \lambda, \regSet_{\rm S}, \PRmatSet, \mu,
\regSet_{\dpsi})$, where $\lambda$ and $\regSet_{\rm S}$ are, respectively, the
strength and form of regularization for $\srVec$.  We would also
obtain the evidence by marginalizing both the source-intensity and the
potential correction values, $P(\dataVec|\responseSet, \lambda,
\regSet_{\rm S}, \PRmatSet, \mu, \regSetdpsi) = \int \rm{d}\srVec\,
\rm{d}\dpsiVec\, P(\dataVec|\srVec, \dpsiVec, \responseSet, \PRmatSet)
P(\srVec, \dpsiVec | \lambda, \regSet_{\rm S}, \mu, \regSetdpsi)$.
However, due to the iterative nature of the method (i.e., $\srVec$
and $\dpsiVec$ are not inferred simultaneously), we do not have such
expressions for the posterior and the evidence.  Pragmatically, we use
the evidence from the source reconstruction (given the corrections
$\dpsiVec$), $P(\dataVec | \responseSet, \dpsiVec, \lambda,
\regSet_{\rm S})$, for comparing the potential models, PSF and
regularizations. Specifically, after iterating through the source-intensity 
reconstructions and lens potential corrections, we use the
final corrected lens potential for one last source-intensity
reconstruction and use the evidence from this final source
reconstruction for comparing models.  This approximation is valid
provided that the probability distributions of $\dpsiVec$ and the
regularization constant are sharply peaked at the most probable
values.  \citet{SuyuEtal06} showed that the delta function
approximation for the regularization constant is acceptable;
simulations of the iterative potential reconstruction method suggest
that the probability of $\dpsiVec$ after the final iteration is
sharply peaked.  Therefore, the probability of a given potential
model, PSF, and form of regularization is $P(\responseSet,
\regSet_{\rm S}| \dataVec) \propto \int \rm{d}\lambda\, \rm{d}\dpsiVec
\, P(\dataVec| \responseSet,\dpsiVec, \lambda, \regSet_{\rm S})
P(\responseSet, \regSet_{\rm S}) \sim P(\dataVec|
\responseSet,\hat{\dpsiVec}, \hat{\lambda}, \regSet_{\rm S})
P(\responseSet, \regSet_{\rm S})$, where $\hat{\dpsiVec}$ and
$\hat{\lambda}$ are the most probable solutions. Assuming that all
models are equally probable a priori (i.e., $P(\responseSet,
\regSet_{\rm S})$ is constant), the evidence from the source
reconstruction serves as a reasonable proxy to use for model
comparison.

There is an uncertainty associated with the evidence values due to
finite source-intensity resolution as a result of the source
pixelization.  The source reconstruction region is initially chosen
such that the mapped source region on the image plane encloses the
Einstein ring.  This ensures that the source region contains the
entire source-intensity distribution.  Throughout the iterative
pixelated potential reconstruction, the source region and pixelization
are kept the same.  In the final source reconstruction for evidence
computation, the evidence value depends on the pixelated source region
because the goodness of fit on the image plane generally changes,
especially in areas of significant intensity gradients, as one shifts
the source region.  To estimate the uncertainty in the evidence
values, we perform the last source reconstruction for various source
regions that are shifted by a fraction of a pixel from the optimized
one in the potential reconstruction.  The range of the resulting
evidence values for the various source regions then allow us to
quantify the uncertainty in the evidence.  In addition to the
uncertainty due to source pixelization, the evidence also depends on
the amount of regularization on $\dpsiVec$, which is discussed in 
Section \ref{sec:PPRMethod:matrix:technicalities}.

\subsubsection{Technicalities of the pixelated potential reconstruction}
\label{sec:PPRMethod:matrix:technicalities}

Solving for the potential perturbations is very similar to solving for the
source-intensity distribution in \citet{SuyuEtal06} except for the following
technical details:
\begin{enumerate}
\item In each iteration, the perturbative potential correction is
  obtained only in an annular region instead of over the entire lens
  potential grid due to the need for the source-intensity gradient
  (see Equation (\ref{eq:pertEq})) to be measurable.  Since the
  extended source intensity is only non-negligible near the Einstein
  ring, we only have information about the source-intensity gradients
  in this region.  In practice, the annular region is the mapping of
  the finite source reconstruction grid that encloses the extended
  source with a minimal number of source pixels (for computational
  efficiency).  The annulus of potential corrections obtained at each
  iteration is extrapolated for the next iteration by minimizing the
  curvature in the potential corrections.  This allows the shape of
  the annular region to change as needed when the lens potential gets
  corrected.  In addition, the forms of the regularization matrix, as
  discussed in Appendix A of \citet{SuyuEtal06}, are modified
  accordingly to take into account the nonrectangular reconstruction
  region (described in more detail in the third point below).
\item Since Equation (\ref{eq:djWithPert}) is a perturbative equation
  in $\dpsiVec$, the inversion needs to be \textit{over-regularized}
  to enforce a small correction in each iteration.  Empirically, we
  set the regularization constant, $\mu$, at roughly the peak of the
  function $\mu E_{\rm{\delta \psi}}$ (within a factor of 10), which
  corresponds to the value before which the prior dominates.  The
  resulting evidence value from the final source-intensity
  reconstruction weakly depends on the value of $\mu$, and we include
  this dependence in the uncertainty of the evidence value.
\item The potential corrections are generally nonzero at the edge of
  the annular reconstruction region.  This calls for slightly
  different structures of regularization compared to those written in
  Appendix A in \citet{SuyuEtal06} for source-intensity reconstruction
  (since the source grids are chosen to enclose the entire extended
  source such that edge pixels have nearly zero intensities).  The
  regularizations are still based on derivatives of $\dpsiVec$;
  however, no patching with lower derivatives should be used for the
  edge pixels because the zeroth-order regularization at the top/right
  edge will incorrectly enforce the $\dpsiVec$ values to zero in those
  areas.  The absence of the lower derivative patches leads to a
  singular regularization matrix,\footnote{Having a singular
    regularization matrix ($\hessS$) does not prevent one from
    calculating $\dpsiVec$ because the matrix for inversion
    ($\hessM=\hessD+\mu\hessS$) is, in general, nonsingular.} which is
  problematic for evaluating the Bayesian evidence for lens potential
  correction.  However, since we do not use the evidence values to
  compare the forms of regularization for the potential corrections
  (because we use only the curvature form) nor to compare the lens
  potential and PSF model, the revised structure of regularization is
  acceptable.  We have found this structure of regularization for
  potential corrections to work for various types of sources (with
  varying sizes, shapes, number of components, etc.).
\end{enumerate}

In the source reconstruction steps of this iterative scheme, we
discover by using simulated data that over-regularizing the source
reconstruction in early iterations helps the process to converge.
This is because initial guess potentials that are significantly
perturbed from the true potential often lead to highly discontinuous
source distributions when optimally regularized (corresponding to
maximal Bayesian evidence, which balances the goodness of fit and the
prior), and over-regularization would give a more regularized source-intensity
gradient for the potential correction.  Unfortunately, we do
not have an objective way of setting this over-regularization factor
for the source reconstruction.  Currently, at each source
reconstruction iteration, we set the over-regularization factor such
that the magnitude of the intensity deficit is at approximately the
same level as that from the optimally-regularized case but with a
smoother source-intensity distribution for numerical derivatives.
This scheme ensures that we do not over-regularize when we are close
to the true potential.  Based on simulated test runs, the recovery of
the true potential depends on the amount of over-regularization.  When
the initial guess is far from the true potential, over-regularization
in the early iterations is crucial for convergence.  We find that it
is better to over-regularize in excess than in deficit.  Too much
over-regularization simply leads to more iterations to converge,
whereas too little over-regularization may not converge at all.

For each iteration of source-intensity reconstruction, there is also a
mask on the source plane to exclude source pixels that either (1) are
not mapped by that iteration's lens potential on the data grid or (2)
have no neighboring pixels for the computation of numerical
derivatives.  We generalize the regularizing function for this
nonrectangular reconstruction region to have the right-most and
top-most pixels (pixels adjacent to the edge or adjacent to the masked
source pixels) patched with lower derivatives as we did for the edge
pixels in Appendix A of \citet{SuyuEtal06}.  This patching ensures
that the regularization matrix is nonsingular for the evaluation of
the Bayesian evidence.

Based on simulated test runs, we find that a practical stopping
criterion for the iterative procedure is to terminate when the
relative potential corrections between all image pairs are $(\dpsi_1
- \dpsi_2) / (\psi_1-\psi_2) < 0.1\%$, where $1$ and $2$ label the
images in any pair.  After this criterion is reached, further
iterations give a negligible contribution to the predicted Fermat
potential differences between the images.

\subsubsection{Mass-sheet degeneracy}
\label{sec:PPRMethod:matrix:MSD}

The restriction to using only isophotal data implies that the
potential correction we obtain at each iteration may be affected by
the ``mass-sheet degeneracy'' \citep{FalcoEtal85}. However, the
addition of mass sheets is suppressed by the curvature form of the
regularization for the potential correction and also by the large
amount of over-regularization.  We refer to \citet{KochanekEtal06} and
Paper II for a detailed description of the mass-sheet degeneracy; here
we review a few key points that are relevant for the potential
corrections.  In essence, an arbitrary symmetric paraboloid, gradient
sheet, and constant can be added to the potential without changing the
predicted lensed image:
\be
\label{eq:MassSheetTrans}
\psi_{\nu}(\vec{\theta}) = \frac{1-\nu}{2} \vert \vec{\theta} \vert
^2 + \vec{a}\cdot\vec{\theta} + c + \nu\psi(\vec{\theta}), 
\ee 
where $\psi(\vec{\theta})$ is the original potential,
$\psi_{\nu}(\vec{\theta}) $ is the transformed potential, and $\nu$,
$\vec{a}$, and $c$ are constants.  The constants $\vec{a}$ and ${c}$
have no physical effects on the lens systems as they merely change the
origin on the source plane (which is unknowable) and change the
zero point of the potential (which is not observable).  The parameter
$1-\nu$ refers to the amount of mass sheet, which can be seen in the
corresponding convergence transformation:
$\kappa_{\nu}(\vec{\theta})=(1-\nu)+\nu\kappa(\vec{\theta})$.  To make
sure we remain ``close'' to the initial potential model, we set
$\nu=1$ and fix three points in the corrected potential after each
iteration to the corresponding values of the initial potential.
Setting $\nu=1$ ensures that the size of the extended source intensity
remains approximately the same, and the three fixed points allow us to
solve for $\vec{a}$ and $c$ in Equation (\ref{eq:MassSheetTrans}) to
remove irrelevant gradient sheets and constants in the reconstructed
potential.  We choose the three points to be three of the four (top,
left, right, and bottom) locations of the annular reconstruction
region that are midway in thickness between the annular edges.  The
three points are usually chosen to be at places with lower surface
brightness in the ring.  This technique of ``fixing'' the mass-sheet
degeneracy is demonstrated in Section \ref{sec:PPRMethod:matrix:summary} 
using simulated data.

\subsubsection{Summary}
\label{sec:PPRMethod:matrix:summary}

To summarize, the steps for the iterative and perturbative potential
reconstruction scheme via matrices are as follows. (1) Reconstruct
the source-intensity distribution given the initial (or corrected)
lens potential based on \citet{SuyuEtal06}.  (2) Compute the
intensity deficit and the source-intensity gradient.  (3) Solve
Equation (\ref{eq:djWithPert}) for the potential corrections
$\dpsiVec$ in the annulus of reconstruction.  (4) Update the current
potential using Equation (\ref{eq:potmodel}): $\boldsymbol{\psi}_{\rm
  {next \ iteration}} = \boldsymbol{\psi}_{\rm{current \
    iteration}}-\dpsiVec$.  (5) Transform the corrected potential
$\boldsymbol{\psi}_{\rm {next \ iteration}}$ via Equation
(\ref{eq:MassSheetTrans}) so that $\nu=1$ and the transformed
corrected potential has the same values as the initial potential at
the three fixed points. (6) Extrapolate the transformed corrected
potential for the next iteration. (7) Interpolate the transformed
corrected potential onto the resolution of the data grid for the next
iteration's source reconstruction.  (8) Repeat the process using
the extrapolated and finely gridded reconstructed potential, and stop
the process when the relative potential correction between any pair
of images is $<0.1\%$.


\section{Demonstration: potential perturbation due to an invisible mass clump}
\label{sec:PPRMethod:demo}

In the previous section, we have outlined a method of pixelated
potential reconstruction.  In this section, we will demonstrate this
method using simulated data with a lens consisting of two mass
components.

\subsection{Simulated data}
\begin{figure*} \begin{center}
\includegraphics[width=165mm]{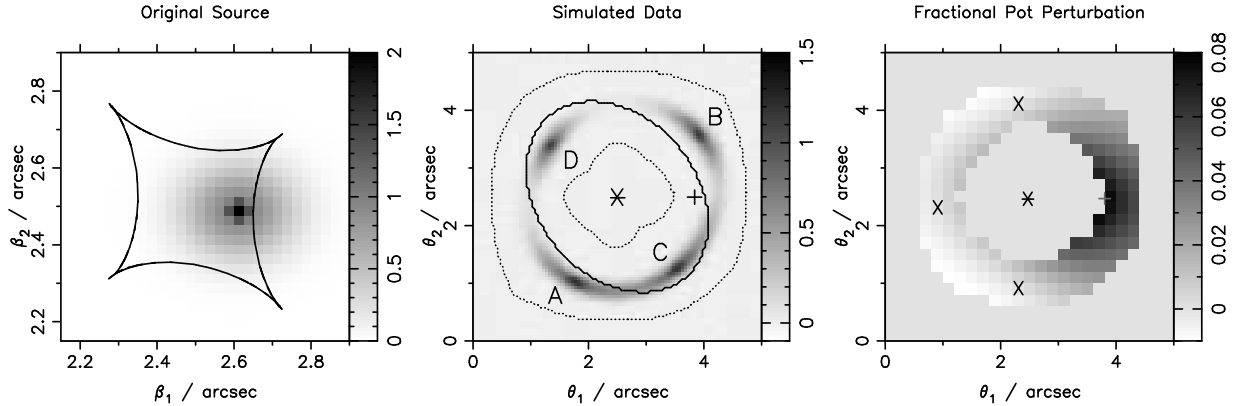}
\end{center} 
\caption[Demonstration of potential reconstruction: simulated data]
{\label{fig:PR:demo1:simData} Demonstration of potential reconstruction:
simulated data and potential perturbation.  Left-hand panel: the simulated
source-intensity distribution with an extended component (of peak intensity of
1.0 in arbitrary units) and a central point source (of intensity 3.0) on a
$30\times30$ grid. The solid curves are the astroid caustics of the initial
potential that consists of only the main SIE.  Middle panel: the simulated
image of the source-intensity distribution on the left using the true potential
consisting of two SIEs (convolution with Gaussian PSF and addition of noise are
included, as described in the text).  The solid line is the critical curve of
the initial potential and the dotted lines mark the annular region to which the
source grid maps (using the mapping matrix $\responseSet$).  Right-hand
panel: the fractional potential perturbation in the initial potential model. 
The Xs mark the three points where we fix the potential perturbation to zero. 
In both the middle and right-hand panels, the asterisk and the plus sign
indicate the positions of the main SIE component and the perturbing SIE
component, respectively.}
\end{figure*}

We use singular isothermal ellipsoid (SIE) potentials
\citep{KormannEtal94} to test the potential reconstruction method.
For this demonstration, we let the lens be comprised of two SIEs at
the same redshift $z_{\rm d}=0.3$: a main component and a perturber.  The
main lens has a one-dimensional velocity dispersion of $260 \rm{\, km \, s^{-1}}$, an axis
ratio of $0.75$, and a semi-major axis position angle of $45^{\circ}$
(from vertical in the counterclockwise direction).  The (arbitrary) origin
of the coordinates is set such that the lens is centered at $(2.5'',
2.5'')$, the center of the $5''\times5''$ image.  The perturbing SIE
is centered at $(3.8'', 2.5'')$ with a velocity dispersion of $50
\rm{\, km\, s^{-1}}$, axis ratio of $0.60$, and semimajor axis position
angle of $70^{\circ}$.  The exact potential is the sum of these two
SIEs.  We model the source intensity as an elliptical distribution
inside the caustics at $z_{\rm s}=3.0$ with an extended component (of peak
intensity of 1.0 in arbitrary units) and a central point source (of
intensity 3.0).  This source is chosen such that the lensed image
resembles B1608+656.  We use $100\times100$ image pixels each of size
$0.05''$ (typical pixel size of the \HST Advanced Camera for Surveys
(ACS)), $30\times30$ source pixels each of size $0.025''$, and
$25\times25$ potential pixels each of size $0.2''$.  To obtain the
simulated data, we map the source-intensity distribution to the image
plane using the exact lens potential and the lens equation, convolve
the lensed image with a Gaussian PSF whose $\rm{FWHM}=0.15''$ and add
Gaussian noise of variance $0.015$.  Fig.~\ref{fig:PR:demo1:simData}
shows the simulated source in the left-hand panel and the simulated
noisy data image in the middle panel.  The Fermat potential difference
between the images are listed in Table \ref{tab:PR:demo1:fermPot}.
The images are labeled by A, B, C, D, and their locations are
$(1.77'',1.02'')$, $(3.90'',3.59'')$, $(3.54'',1.26'')$, and
$(1.34'',3.38'')$, respectively.

\begin{table*}
\begin{center}
\caption[Demonstration of potential reconstruction: actual and predicted Fermat potential differences]{\label{tab:PR:demo1:fermPot} The relative Fermat potential ($\phi = (\vec{\theta}-\vec{\beta})^2/2-\psi$) between the four images of the true potential and of the reconstructed potential for a few selected iterations  }
\begin{tabular}{c c c c c c}\tableline \tableline
 & & & & Source Position \\
Potential  & $\phi_{AB}$ & $\phi_{CB}$ & $\phi_{DB}$ & (arcsec) \\
\tableline
True & 0.141 & 0.234 & 0.437 & $\dots$ \\
Initial &  $0.172\pm0.189$ & $0.228\pm0.156$ & $0.437\pm0.041$ & $(2.587,2.483)\pm(0.013,0.076)$ \\
Iteration=0 & $0.178\pm0.070$ & $0.246\pm0.068$ & $0.479\pm0.010$ & $(2.608, 2.483)\pm(0.006,0.034)$ \\
Iteration=2 & $0.161\pm0.011$ & $0.242\pm0.010$ & $0.471\pm0.011$ & $(2.623, 2.484)\pm(0.005, 0.005)$ \\
Iteration=9 & $0.151\pm0.006$ & $0.244\pm0.004$ & $0.454\pm0.006$ & $(2.621, 2.484)\pm(0.003, 0.002)$ \\
\tableline
$\nu$ & 0.96 & & & \\
Iteration=9 & $0.145\pm0.006$ & $0.234\pm0.004$ & $0.436\pm0.006$ & \\
\tableline
\end{tabular}
\end{center}
\tablecomments{We use the average source position of the four source positions for the computation of the Fermat potential.  The four source positions deviate by $\sim 0.1''$ in the initial model, and agree within $\sim 0.005''$ at iteration=9.  The uncertainties in the predicted relative Fermat potential are due to the uncertainties in the source position.  The good agreement between the predicted Fermat potential values for the initial potential and the true values is coincidental due to the use of the average source position.}
\end{table*}

\subsection{Iterative and perturbative potential corrections}

We take the initial guess of the lens potential to be the main SIE
component but with the position angle changed from $45$ to $40^{\circ}$.  
This corresponds to a typical scenario where the perturbing
SIE is faint/dark so that it is not detected in the image, and hence is
not incorporated in the smooth parametrized model of the main SIE
component.  The rotation in the position angle of the main SIE
component corresponds to a situation where the mass of the galaxy does
not strictly follow the light, but the position angle of the lens mass
distribution is initially adopted from the position angle of the lens
galaxy light.  Here and after, ``initial potential'' refers to this
initial guess of the potential model (as opposed to the true/exact
potential).  Fig.~\ref{fig:PR:demo1:simData} shows the potential
perturbation relative to the initial potential in the right-hand
panel. In obtaining this plot, the initial potential has a constant
gradient plane and offset added such that the top, left, and bottom
midpoints in the annulus (marked by Xs in the plot) are fixed to the
true potential with zero potential perturbation (as described in the
passage following Equation (\ref{eq:MassSheetTrans})).  In the
iterative potential reconstruction process, the reconstructed
potential at each iteration also has these three points in the annulus
fixed to the initial model.  The locations of the three fixed points
have no impact on recovering the true potential when the source is
extended enough to form an Einstein ring on the image plane.  However,
if the source is compact, then locations of the three points do matter
and they are chosen to be at places where the information content
(image intensity) is low.

We perform 10 iterations of the perturbative potential correction
method outlined in Section \ref{sec:PPRMethod:matrix}.  The iterations
are labeled ``PI'' from 0 to 9.  For each source reconstruction
iteration, we adopt the curvature form of regularization and use the
source-intensity reconstruction for the evaluation of the source-intensity 
gradients that are needed for the potential correction.  The
source inversions are over-regularized in early iterations in order to obtain
smooth source reconstructions for evaluating the gradients.  For each
potential correction iteration, we use the curvature form of
regularization, and set the regularization constant for the potential
reconstruction to be $10\times$ the value of $\mu$ where $\mu
E_{\rm{\dpsi}}$ peaks in iteration=0.  This regularization value is
$\sim 10^8$ and is used for all subsequent iterations (since we find
that the peak in $\mu E_{\rm{\dpsi}}$ changes little as the iterations
proceed).  For comparison, the ``optimal'' regularization constant is
$\sim 10^2$ at iteration=0 and is $\sim 10^7$ at iteration=9.
Therefore, the potential reconstruction inversions are heavily
over-regularized in the early iterations to keep the corrections to
first order; as the lens potential gets corrected, the amount of
over-regularization diminishes as the inversion approaches the linear
regime with small intensity deficits.  We show figures of source
reconstructions and potential corrections for some, but not all, of
the iterations.

\begin{figure*}
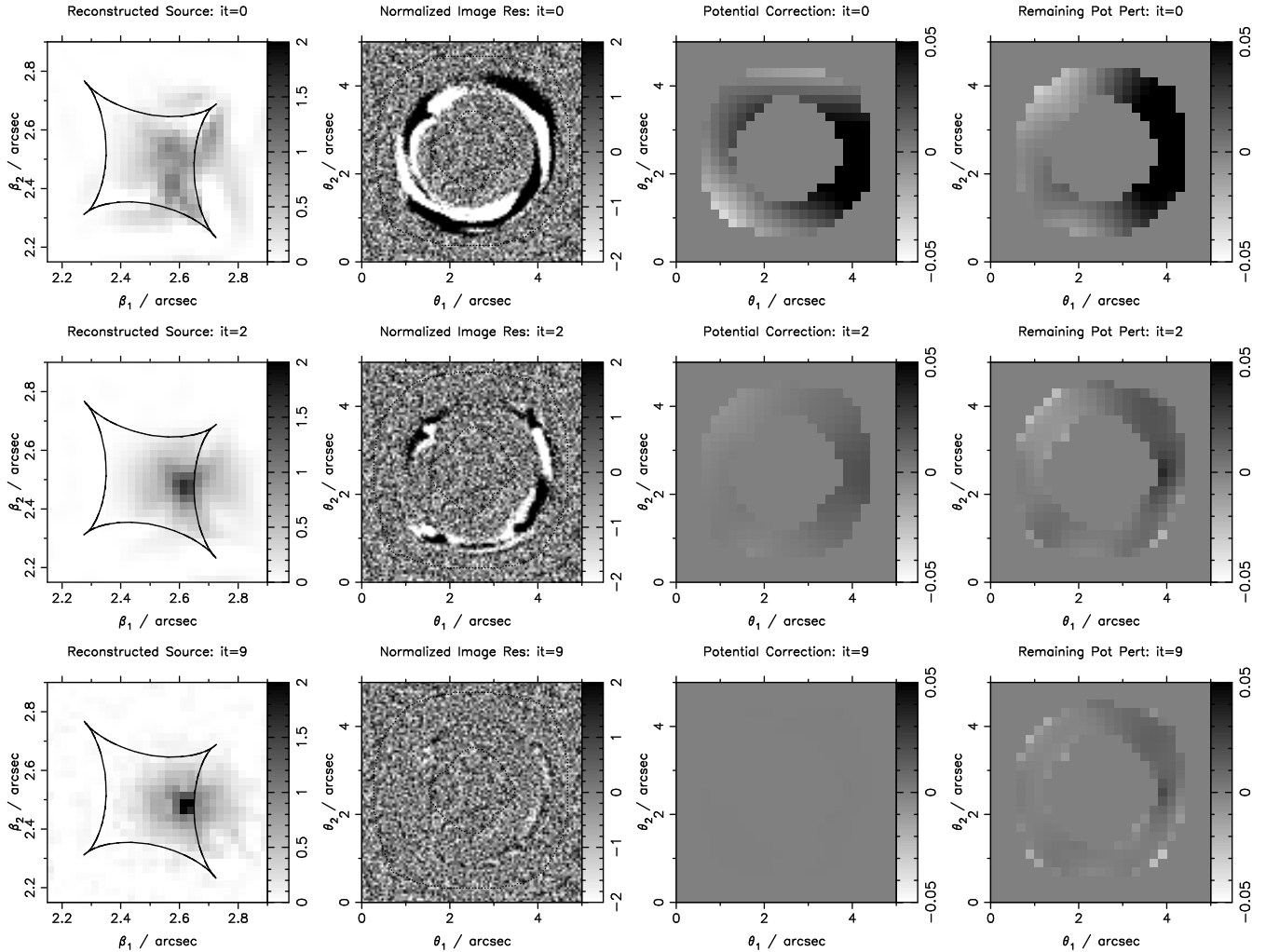

\begin{center}
\includegraphics[width=180mm]{f2a.ps}
\includegraphics[width=180mm]{f2b.ps}
\includegraphics[width=180mm]{f2c.ps}
\end{center}
\caption[Demonstration of potential reconstruction: results of source-intensity 
reconstruction and potential correction for iteration = 0, 2,
and 9] {\label{fig:PR:demo1:itn0_2_9} Demonstration of potential
  reconstruction: results of source-intensity reconstruction and
  potential correction for iteration = 0, 2, and 9.  The top row shows
  the results for PI=0.  Left-hand panel: the reconstructed source
  intensity using curvature regularization that is over-regularized to
  ensure a smooth resulting source for evaluation of the gradients.
  The caustic curves in solid are those of the initial potential.
  Middle-left panel: the normalized image residual (difference between
  the simulated image and the predicted image from the reconstructed
  source in the left-hand panel, in units of the estimated pixel
  uncertainty from the data image covariance matrix).  The prominent
  arc features are due to the potential perturbation.  Middle-right
  panel: the reconstructed $\dpsiVec$ using the source-intensity
  gradients and image residual.  Right-hand panel: the amount of
  potential perturbation that remains to be corrected.  The middle and
  bottom rows show the results for PI=2 and PI=9, respectively, with
  the panels arranged in the same way as in the top row.  As the
  iterative potential correction proceeds, the source resembles better
  the original source in Fig.~\ref{fig:PR:demo1:simData}, the image
  residual becomes less prominent, and the magnitude of the
  reconstructed $\dpsiVec$ decreases.  At PI=9, the source in the
  left-hand panel has been faithfully reconstructed that results in
  negligible image residual in the middle-left panel.  The remaining
  potential perturbation in the right-hand panel, now close to zero,
  cannot be fully corrected due to the noise in the data.}
\end{figure*}

The top row of Fig.~\ref{fig:PR:demo1:itn0_2_9} shows the results of
PI=0.  The over-regularized reconstructed source in the left-hand
panel does not resemble the original source, and the (normalized)
image residual in the middle-left panel shows prominent arc features
due to the presence of both the misaligned initial model and the SIE
potential perturbation.  The reconstructed $\dpsiVec$ in the
middle-right panel is of the same structures as the exact $\dpsiVec$
in Fig.~\ref{fig:PR:demo1:simData}, though the magnitude is smaller
due to the correction being a perturbative one.  A plot of the image
residual after correction $(=\dIVec - \PRmatSet\dpsiVec)$ continues to
show arc features though less prominent than in the top middle-left
panel in Fig.~\ref{fig:PR:demo1:itn0_2_9}.  The same image residual
plot with the true potential perturbation also shows similar arc
features, which indicates that Equation (\ref{eq:pertEq}) is indeed a
perturbative equation and thus justifies the over-regularization in
the potential correction step.

The second row of Fig.~\ref{fig:PR:demo1:itn0_2_9} shows the results
of PI=2.  The reconstructed source in the left-hand panel better
resembles the original source in Fig.~\ref{fig:PR:demo1:simData}.  The
amount of misfit in the image residual has decreased in the
middle-left panel, signaling that we are correcting toward the true
potential.  The middle-right panel is the potential correction in
PI=2, and the right-hand panel is the amount of perturbation that
remains after PI=2.  The amount of potential perturbation remaining is
closer to zero compared to the top row, which is a sign that the
iterative method converges.

The bottom panels in Fig.~\ref{fig:PR:demo1:itn0_2_9} show the results
of PI=9, the last iteration.  The source is faithfully recovered in
the left-hand panel, resulting in negligible image residual in the
middle-left panel (reduced $\chi^2=1.02$ inside the
annulus\footnote{The reduced $\chi^2$ is given by $\chi^2 /(
  N_{\rm{pix\ in\ annulus}}-\gamma)$, where $N_{\rm{pix\ in\
      annulus}}$ is the number of data pixels in the annulus that
  encloses the ring and $\gamma$ is an estimate of the number of
  ``effective'' parameters \citep[e.g.][]{MacKay92, SuyuEtal06}.}).
The centroid of the source is slightly shifted compared to the
original because of our adding constant gradients to fix the three
points in the potential corrections.  The absolute position of the
source is irrelevant as we can arbitrarily set the coordinates; it is
only the relative positions on the source plane that matter.  The
source positions are shifted \textit{relative} to the plotted caustic
curve only because these caustic curves are the ones from the initial
potential guess (they were not computed for the reconstructed
potential due to the low resolution in the reconstructed potential
grids).  If we were to plot the caustic curve of the corrected
potential, we would find no overall shift in the source with respect
to the caustic curve.  The middle-right panel shows the final
iteration's potential correction, which is barely visible due to the
negligible image residual left to correct.  The right-hand panel shows
that most of the potential perturbation to the true potential has been
corrected, though there is still some left.  However, this amount of
remaining uncorrected potential perturbation leads to image residuals
that are effectively masked by the noise in the data.  We have thus
reached the limit in the potential correction that is set by noise in
the data.

Table \ref{tab:PR:demo1:fermPot} lists the predicted Fermat potential
differences for the initial potential guess and for the corrected
potential in PI=0, 2, and 9.  We use the average source position (also
listed in the table) of the four mapped source positions for the
computation of the Fermat potential.  The uncertainty in the predicted
Fermat potential difference comes from the error in the source
position due to discrepancies in the mapped source positions of the
four images.  The mapped source positions agree within $\sim 0.005''$
(i.e., within a fifth of a source pixel) in the final iteration, a
significant improvement to $\sim 0.1''$ in the initial potential.  The
convergent Fermat potential differences in PI=9 are systematically
higher than the true Fermat potential differences.  This is because
lensing only allows us to recover the Fermat potential differences up
to a constant factor due to the mass-sheet degeneracy.  The
transformation in Equation (\ref{eq:MassSheetTrans}) would scale the
Fermat potential difference by a factor of $\nu$.  The last row in
Table \ref{tab:PR:demo1:fermPot} shows that a mass-sheet
transformation with $\nu=0.96$ leads to the predicted Fermat potential
values agreeing with the true values within the uncertainties.  We
expect this particular simulation's reconstructed pixelated
potential to be different from the true potential by a mass-sheet
transformation of $\nu\sim 0.96$ due to the unaccounted mass of the
secondary SIE ($\sim 4\%$ of the primary SIE) in the initial model.
In the iterative potential corrections, mass additions are suppressed
in the annulus due to the regularization. This breaks the mass-sheet
degeneracy, but underestimates the total mass within the annulus (the
SIE perturber was not included in the initial model): the
reconstructed potential, therefore, continues to have a deficit of mass
in the annulus.  Since the value of the convergence in the annulus is
generally less than $1$, the reconstructed potential is thus approximately a
mass-sheet transformation of the true potential with the mass deficit
in the form of a constant sheet.
  
The simulation we have shown is one of the worst-case scenarios where
even the total mass of the initial lens model enclosed within the Einstein
ring is wrong.  For initial potential models that have the correct
amount of mass within the Einstein ring (this enclosed mass is what
lensing can robustly measure to $\sim 1\%-2\%$ accuracy in real systems)
and with the mass-sheet degeneracy broken (using external information
such as stellar dynamics), the reconstructed potential would
faithfully recover the Fermat potential.

\subsection{Discussion}

This demonstration shows that the iterative and perturbative potential
reconstruction method works in practice. Using simulated data, we find
that potential perturbations $\lesssim 5\%$ (which may correspond to
as much as $\sim 20\%$ in the relative potential perturbations between
image pairs) are correctable, though the actual amount depends on the
amount of over-regularization for both the source inversion and the
potential correction, and on the extendedness of the source-intensity
distribution.  In the case where the solution converges, the
magnitudes of the relative potential corrections between image pairs
steadily decrease, and we end the iterative procedure when the
stopping criterion (described in Section \ref{sec:PPRMethod:matrix})
is met.

Regarding the size of the source-intensity distribution, the more
extended a source is, the better we can recover the potential.  When
the source is extended enough to be lensed into a closed ring, the
true potential can be fully recovered (up to the limit set by the
noise in the data) from potential corrections based on Equation 
(\ref{eq:djWithPert}).  When the source is extended to cover about
half of the Einstein ring, then the corrected potential faithfully
reproduces the source with negligible image residual, but the relative
Fermat potentials may not be recovered due to a slight relative offset
in the potential between the images.  This is because the ``connecting
characteristics'' (see \citet{SuyuBlandford06}) that fix the potential
difference between the images go through regions without much
signal (light of the lensed source).  Therefore, the potential is
locally corrected at regions near the images (where there is light),
but the global offset between the regions cannot be determined.

For sources that are small in extent, the potential correction also
depends on the points we choose to fix to the initial potential model.
Since an isolated image is generally more prone to having its
potential be offset relative to the other images, we set two of the
three fixed points in the gaps on both sides of the most isolated
image and one point near the connecting images.

We find that a wrong PSF model (e.g., of a different width) would lead
to intensity deficit that would not be correctable by the iterative
potential reconstruction method.  Therefore, an uncorrectable image
residual is a sign that our model of the system (other than the lens
potential) is wrong.

The potential grid that we used was $25\times25$, which we find to be
a good balance between the number of degrees of freedom and goodness
of fit.  The higher the number of potential pixels, the better one can
fit to the image residual; however, in this case, it is also more
probable to have degenerate solutions.  The Bayesian evidence from the
source reconstruction in principle can be used to compare the
different potential grids.  In general, we find that a potential grid
that is $\sim 4$ times coarser than the image grid works well.

In Section \ref{sec:PPRMethod:realData}, we generalize this iterative potential
reconstruction method, which has been shown to work on simulated data,
to treat real gravitational lens images such as B1608+656.


\section{Generalization to realistic data: incorporating dust
  extinction and lens galaxy light}
\label{sec:PPRMethod:realData}
In the previous section, we have demonstrated the method of
pixelated potential reconstruction using simulated data.  In the mock
data, only the image of lensed source was there; in reality, there
would also be light from the lens galaxy.  Furthermore, in some cases,
such as B1608+656, dust is present and absorbs light from both the
source galaxy and the lens galaxy.  Based on results of the previous
section, an accurate extraction of the light from the lensed extended
source is crucial for reconstructing the lens potential.  Therefore,
we will generalize the formalism given in Section \ref{sec:PPRMethod}
to incorporate the lens galaxy light and dust.

Suppose that we have a set of PSF, dust, and lens galaxy light
models (the process of obtaining these models is described in detail
in Section \ref{sec:ImProc}), a lens potential model, and
the observed image.  Separating the observed image into two
components, the lensed source and the lens galaxy, we can
model the observed image (as a vector for the intensities of the
image pixels) as 
\be
\label{eq:dataVecComp}
\dataVec = \overbrace{\blurSet \cdot \dustSet \cdot \lensSet \cdot
  \srVec}^{\rm{lensed \ extended \ source}} + \overbrace{ \blurSet
  \cdot \dustSet \cdot \glightVec}^{\rm{lens\ galaxy}} +\ \noiseVec,
\ee 
where $\blurSet$ is a PSF blurring matrix, $\dustSet$ is a dust
extinction matrix, $\lensSet$ is the lensing matrix (containing the
lens potential model), $\srVec$ is the source-intensity distribution,
$\glightVec$ is the lens galaxy intensity distribution, and
$\noiseVec$ is the noise in the data characterized by the covariance
matrix $\imCM$.  This is an extended version of the equation
$\dataVec=\responseSet\srVec+\noiseVec$  in \citet{SuyuEtal06} with
$\responseSet$ replaced by $\blurSet \cdot \dustSet \cdot \lensSet$
and $\dataVec$ replaced by $\dataVec -
\blurSet\cdot\dustSet\cdot\glightVec$.  The order of the matrix
products in both terms are obtained by tracing backwards along the
light rays: we first encounter the PSF blurring from the telescope
($\blurSet$), then dust extinction ($\dustSet$) in the lens plane,
then the strong lensing effects ($\lensSet$) in the case of the lensed
source, and finally the origin of light ($\srVec$ or $\glightVec$).

Here we assume that the dust lies in a screen in front of the lensed
source and the lens galaxy.  This assumption is not strictly valid for
the lens galaxy if the dust were to have originated from G2
\citep{SurpiBlandford03}.  In this case, the dust and stars are
mingled together in the lens galaxy.  It is beyond the scope of this
paper to treat this mixed light and dust problem.  However, we note
that the dust screen assumption is acceptable since the aim is to
obtain an accurate lensed source-intensity distribution (for which the
dust screen assumption is valid) and not the lens galaxy
intensity distribution near the core where the mixing effects would
dominate.  Furthermore, in simple toy models, where either the dust and
stars are uniformly mixed or the dust is a screen lying inside
the lens galaxy, we find that the extinction of the lens galaxy light
is well approximated as extinction by a foreground dust screen with a
reduced visual extinction.  Our simple foreground dust screen model
thus provides an {\it effective} extinction that incorporates the
reduced extinction for the lens and the full extinction by a
foreground dust screen for the lensed source.

If the lensed source contains a bright core such as an active galactic nucleus (AGN), then we
could consider extending Equation (\ref{eq:dataVecComp}) and model the observed
image as
\be
\label{eq:dataVecCompAGN}
\dataVec = \blurSet \cdot \dustSet \cdot \lensSet \cdot \srVec +
\sum_{i=1}^{N_{\rm{images}}} \dust_i \alpha_i {\rm PSF}(\vec{\theta}_i) + 
\blurSet \cdot \dustSet \cdot \glightVec + \noiseVec, 
\ee
where the light from the extended part of the host (the first term) would be
modeled
separately from that from 
the point sources (the second term), and $\alpha_i$ are the
intensities (flux per unit solid angle in a pixel) of the point sources 
(which are generally not the
same for all images due to finite resolution---both lensing and microlensing
give rise to different magnification of the point-like source---and, 
in the case of a time-varying core, time delay difference).  
However, it is the extended image surface brightness that provides
the information needed
to reconstruct the lens potential.
For B1608+656, by taking into account
the errors in the modeling associated with the presence of the point sources
(see Section \ref{sec:ACSimprocess}), we will find that a separate modeling of
the point sources is not necessary for reconstructing the lens potential.

Given $\blurSet$, $\dustSet$, $\glightVec$, $\lensSet$ and $\dataVec$,
one can solve for the most probable source-intensity distribution
$\srMPVec$, as in \citet{SuyuEtal06}.  Furthermore, one can use the
Bayesian evidence of the source reconstruction to rank different
models of PSF, dust extinction, lens galaxy light, and lens potential
(see Section \ref{sec:PPRMethod:matrix:probTheory}).  When we compare
models, we mark an annular region enclosing the Einstein ring and use
the same annulus of data for all models (where models refer
collectively to the lens potential, PSF, dust, lens galaxy light, and
regularization).  For the chosen data set, we determine the source
region that maps to the annular region and reconstructs the source
intensities in this region.  The shape of this source region is
generally not rectangular, so we generalize the regularization schemes
in Appendix A of \citet{SuyuEtal06} to patch the right-most and
top-most pixels (pixels adjacent to the edge of grid or adjacent to
the unmapped source pixels) with lower derivatives.  We will use the
Bayesian evidence values from the source reconstruction in Sections
\ref{sec:ImProc} and \ref{sec:PotRec:B1608} to compare various PSF,
dust, lens galaxy light and lens potential models for B1608+656.

To include the effects of galaxy light and dust in the pixelated
potential reconstruction method, we incorporate $\dustSet$ and
$\glightVec$ into Equation (\ref{eq:djWithPert}) as in Equation
(\ref{eq:dataVecComp}), and include $\dustSet$ into $\PRmatSet$ (see
the Appendix for this inclusion).  After these adjustments, we can
iteratively correct for the lens potential in real systems given a
PSF, a dust, and a lens galaxy light model based on the machinery we
developed in the previous sections.

To conclude, we have outlined and demonstrated an iterative and
perturbative potential correction scheme where the accuracy in the
reconstruction is limited by the noise in the data.  The inputs for
this method are an initial guess of the lens potential as well as
assumptions regarding the PSF, dust, and lens galaxy light.  The
outputs are the reconstructed potential on a grid of pixels, the
reconstructed source-intensity distribution, and the Bayesian evidence
from source reconstruction, given the assumptions.  Our goal is to
apply this method to the well-observed lens system B1608+656, and we
begin by describing our \HST observations
of B1608+656 in Section \ref{sec:ImProc}.

\section{Image processing of B1608+656}
\label{sec:ImProc}

\subsection{\HST observations of B1608+656}
\label{sec:ImProc:ACSobs}

B1608+656 was observed with the ACS camera on \HST in the F606W and
F814W filters in 2004 August (Proposal 10158; PI:Fassnacht),
specifically to get high signal-to-noise ratio (S/N) images of the lensed
source emission.  Table \ref{tab:B1608HSTobs} summarizes the
observations.  Each orbit of the ACS visits consisted of one
four-exposure dither pattern in either F606W or F814W through the Wide
Field Channel (WFC).  We used the same dither pattern described in
\citet{YorkEtal05} to permit drizzling to a higher angular resolution
than the default ACS CCD pixel size ($\sim 0.05''$).  This subpixel
scale is especially important for characterizing the PSF.

In order to correct for the dust extinction in the lens system, we also include
the Near Infrared Camera and Multi-Object Spectrometer 1 (NICMOS) F160W images
(Proposal 7422; PI:Readhead).  Details of the NICMOS observations are also
listed in Table \ref{tab:B1608HSTobs}.

\begin{table*}
\begin{center}
\caption{\label{tab:B1608HSTobs} \HST observations of B1608+656}
\begin{tabular}{ccccccc}
\tableline
\tableline
Proposal & Proposal & Date & Instrument & Filter & Exposures & Exposure Time \\
PI & ID & & & & & (s) \\
\tableline
C. Fassnacht &  10158 & 2004 Aug 24 & ACS/WFC & F606W & 4 & 609 \\
& & & & & 4 & 646 \\
& & & & F814W & 4 & 632 \\
& & & &  & 4 & 646 \\
& & 2004 Aug 25 & ACS/WFC & F606W & 8 & 609 \\
& & & & & 8 & 646 \\
& & & & F814W & 8 & 632 \\
& & & &  & 8 & 646 \\
& & 2004 Aug 29 & ACS/WFC & F606W & 4 & 609 \\
& & & & & 4 & 646 \\
& & & & F814W & 4 & 632 \\
& & & &  & 4 & 646 \\
& & 2004 Sept 17 & ACS/WFC & F606W & 4 & 609 \\
& & & & F814W & 4 & 632 \\
& & & &  & 4 & 646 \\
& & & &  & 4 & 646 \\
A. Readhead & 7422 & 1998 Feb 7 & NIC1 & F160W & 5 & 3840 \\
& & & & & 1 & 2048 \\
& & & & & 1 & 896 \\
\tableline
\end{tabular}
\end{center}
\end{table*}

The ACS images of B1608+656 are presented in
Fig.~\ref{fig:B1608acsF606F814} and show the two lensing galaxies and
the presence of a dust lane through the system.  We need to correct
for both the dust lane and the light from the lens galaxies, which can
affect the isophotes of the Einstein ring of the extended lensed
source.  Before we can determine the amount of extinction, we need to
first unify the resolutions of the images in different wavelength
bands due to PSF dependencies.  This requires PSF modeling,
deconvolution, and reconvolution for images.  Having unified the
resolutions of the images, we can determine the intrinsic colors of
the various components (lens galaxies, lensed source galaxy, AGN at
the core of the source galaxy) in the system that are required for the
dust correction.  After correcting for dust, we can then determine the
light profiles of G1 and G2 by fitting them with S\'ersic profiles
($I(r) \propto \exp(-(r/a)^{1/n})$ where $r$ is the radial coordinate,
$a$ is a scale length, and $n$ is known as the S\'ersic index;
\citep{Sersic68}).  It is only at this stage, with the PSF, dust map, and
lens galaxies' light profiles, that we can recover the lensed Einstein ring
surface brightness distribution for lens potential modeling.

To execute the above plan of attack, in Section
\ref{sec:drizzling}, we begin by describing the drizzling process for the ACS
images that are used for the analysis.  In Sections \ref{sec:psf}--\ref{sec:lensGalLight},
 we present a suite of PSF, dust, and lens
galaxies' light models and describe in detail how they are obtained.
Finally, in Section \ref{sec:ImProcModelComp}, we compare these
models.


\subsection{Image drizzling}
\label{sec:drizzling}

In the following subsections, we briefly describe the drizzling process for
combining the dithered ACS images and discuss the alignment of the NICMOS image
to the ACS image.


\subsubsection{ACS image processing}
\label{sec:ACSimprocess}

\begin{figure*}
\begin{center}
\includegraphics[width=75mm]{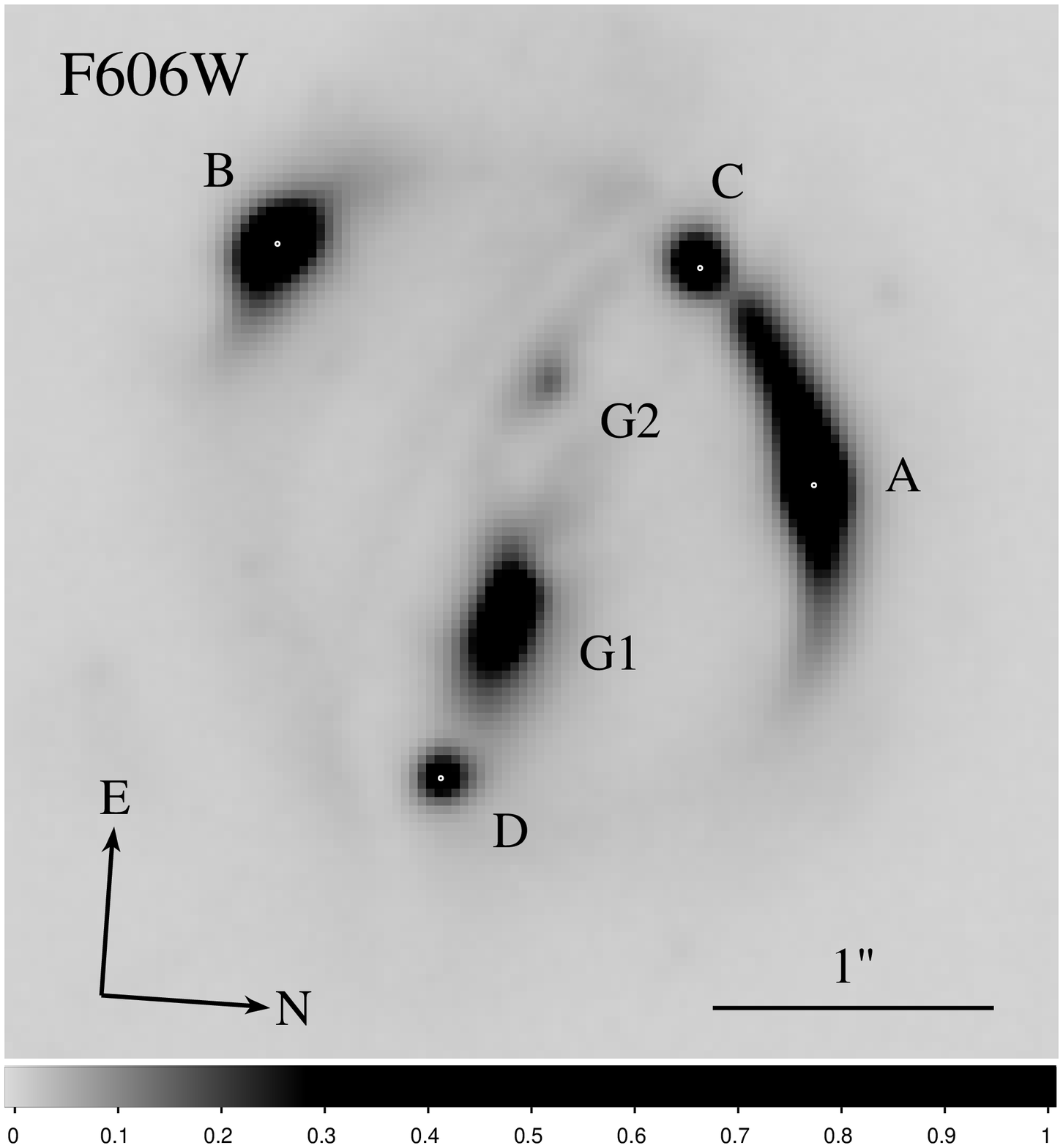}
\includegraphics[width=75mm]{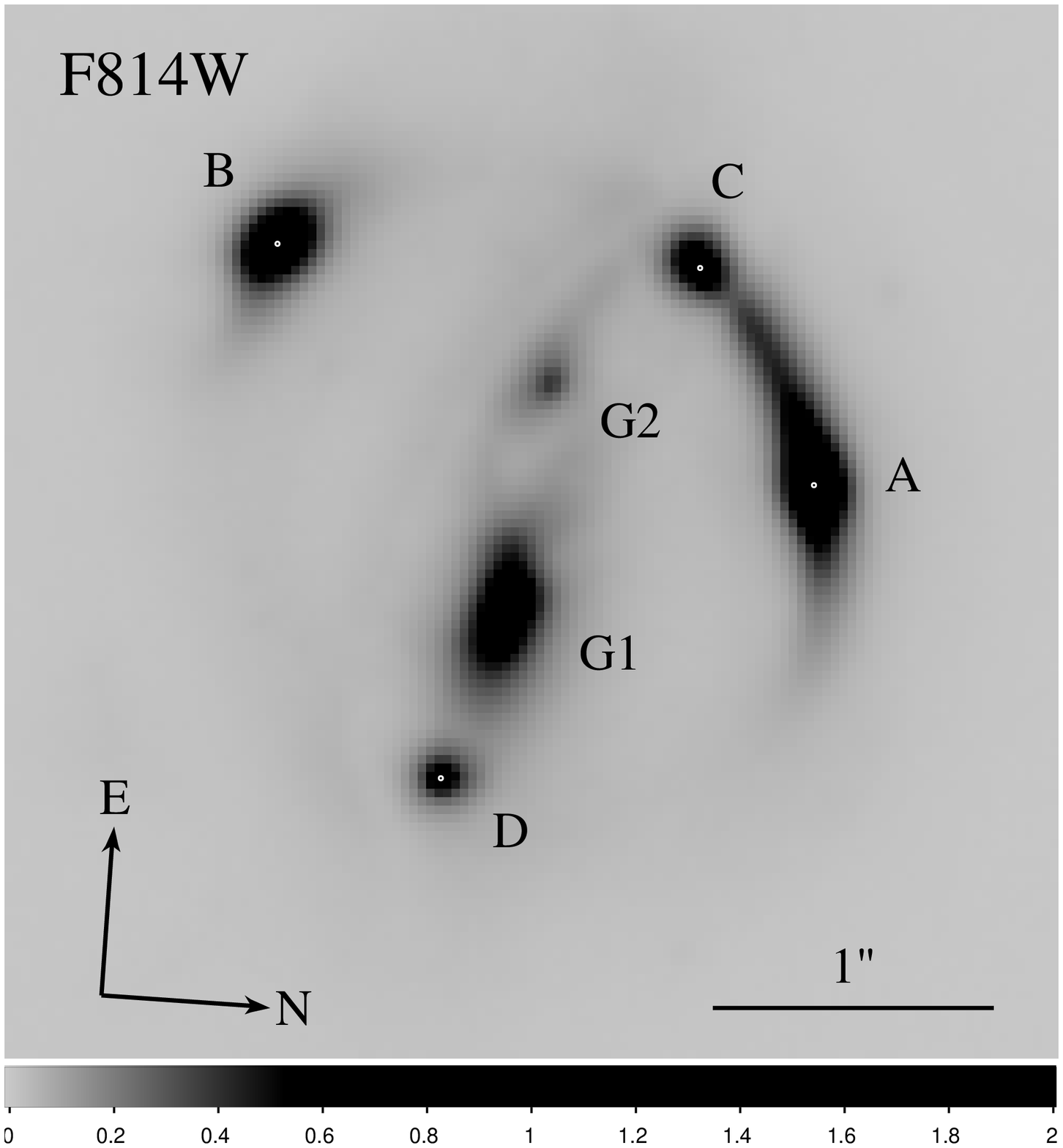}
\caption[Drizzled \HST ACS F606W and F814W images]
{\label{fig:B1608acsF606F814} Left-hand (right-hand) panel: drizzled
  \HST ACS F606W (F814W) images with $0.03''$ pixels from 9 (11) \HST
  orbits.  The dust lane and interacting galaxy lenses are clearly
  visible.  The white dots indicate the centroid positions of the
  images.}
\end{center}
\end{figure*}

The ACS data were reduced using the multidrizzle package
\citep{KoekemoerEtal02} in an early version of the HAGGLeS
image-processing pipeline (P. J. Marshall et al. 2009, in preparation), producing
drizzled images with a $0.03''$ pixel scale.  The drizzled ACS images
are shown in Fig.~\ref{fig:B1608acsF606F814}.  The corresponding
output weight images from multidrizzle give the values for the inverse
variance of each pixel.  We approximate the noise covariance matrix as
diagonal and use the variance pixel values for the diagonal entries,
even though drizzling will correlate the noise between adjacent
pixels.  It is assumed that the effect of drizzling can be modeled as
having a diagonal covariance matrix with the diagonal elements
rescaled \citep{CasertanoEtal00}.  In practice, we do not need to do
the rescaling because the ranking of the models using the
\textit{relative} log evidence values from the source reconstruction
is insensitive to rescaling of the covariance matrix.

A pixelated representation of a continuous intensity distribution
generally introduces error in the interpolated intensity values
between pixels, especially for intensity distributions with sharp
features.  
This error should be 
incorporated into the likelihood function.
Therefore,
for modeling the source-intensity distribution on a grid (in Sections
\ref{sec:ImProcModelComp} and \ref{sec:PotRec:B1608}), we also include
the error due to pixelization on the image and source planes (which we
call ``regridding error'') in the image covariance matrix.  We express
the regridding error on the image plane in terms of the data (instead
of on the source plane and transforming it to the image plane) in
order to obtain a noise map that is independent of the pixelated lens
modeling.  The regridding error associated with pixel $i$ is
\be
\label{eq:regridError}
(\sigma_{\rm grid}^2)_i = \frac{1}{12} \mu_i
\frac{\Delta\beta^2}{\Delta\theta^2}
\sum_{\parbox{17mm}{\centering\scriptsize\it j $\in$ {\rm pixels} \\ {\rm adjacent\ to\ } i}}^{N_{\rm adj}} \frac{(d_j-d_i)^2}{N_{\rm adj}}, 
\ee 
where $\mu_i$ is the lensing magnification at pixel $i$, $\Delta\beta$
is the source pixel size, $\Delta\theta$ is the image pixel size,
$N_{\rm adj}$ is the number of pixels adjacent to pixel $i$, and $d_i$
($d_j$) is the image intensity at pixel $i$ ($j$).  The summation
divided by $\Delta\theta^2$ in the above equation is a conservative
estimate on the error due to pixelization on the image plane. 
Since sharper features in the image have larger gradients (hence,
larger values for the summations), the regridding error is higher in
these areas by construction.
The
remaining quantities in the equation, $\mu_i\Delta\beta^2 / 12$,
account for the uncertainty in the predicted image (the source image
mapped to the image plane) due to the pixelization of the source-intensity 
distribution.  The factor $1/12$ is the second moment of a
uniform distribution between $-0.5$ and $0.5$.  When one constructs
the predicted image by mapping each image pixel to the source plane
and reading off the source-intensity value, the mapped source position
(of an image pixel) is generally not centered on a source pixel, but
have on average a $(1/\sqrt{12})$-pixel shift from the center of the
source pixel.  Therefore, $\Delta\beta/\sqrt{12}$ is the effective
size of the source pixel, which is then magnified by (on average)
$\sqrt{\mu_i}$ due to lensing.  In the pixelated potential
reconstruction, we approximate the magnification at each image pixel
(which requires the second derivative of the potential) by the value
computed from the initial potential because (1) the approximation
enforces the regridding error to be independent of the pixelated
potential modeling and (2) the corrected potential values are
obtained on an annular region only a few pixels thick.  Having
obtained an estimate for the regridding error, we add it in quadrature
to the variance from the weight image to obtain the entries of the
approximated diagonal covariance matrix.  

The inclusion of the
regridding error is important for source-intensity reconstructions
with sharp intensity features (such as the presence of a bright core);
it has the effect of stabilizing the evidence values with respect to 
choices in the source pixelization.  
Without including the regridding error, a pixelated description of,
for example, a source-intensity distribution with a bright core would
be highly sensitive to the centering of the core on the source pixels.
A small mismatch could create large image residuals near the cores
that would veto an otherwise good lensing model, which has the rest of
the extended features well described.  Such an undesirable effect is
mostly removed by the inclusion of the regridding error.
For B1608+656, the ratio of the
regridding error to the error from the multidrizzle weight image is
around $\sim 30$ near the image centroids and $\sim 1$ in other parts
in the Einstein ring.

\subsubsection{NICMOS image processing}
\label{NICMOSimprocess}
\begin{figure}
\begin{center}
\includegraphics[width=80mm]{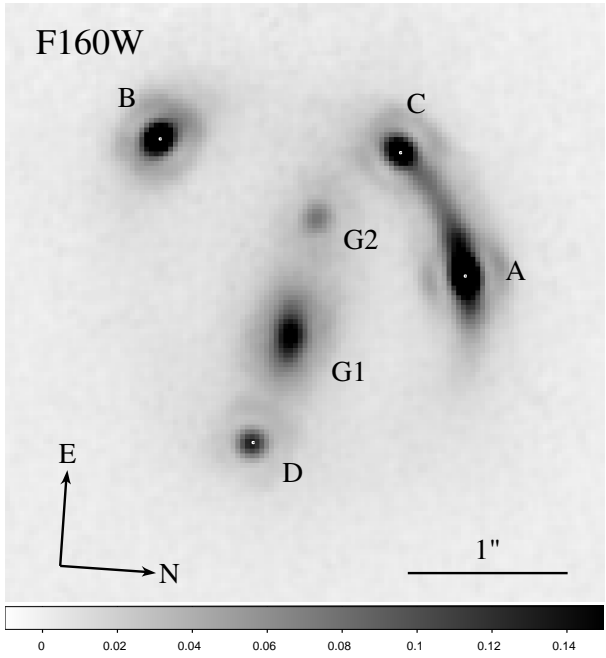}
\end{center}
\caption[SWarped \HST NICMOS F160W image]{\label{fig:B1608nicF160}
  \HST NICMOS F160W image that is SWarped to aligned to the ACS frame
  with a $0.03''$ pixel size.  The white dots indicate the centroid
  positions of the images.}
\end{figure}

The NICMOS F160W image was taken from \citet{KoopmansEtal03}.
Drizzled images on rectangular grids for different instruments are
generally not on the same resolution and not aligned.  This is the
case for the NICMOS and ACS images.  We use SWarp\footnote{A package developed by Emmanuel Bertin at Institut d'Astrophysique de Paris for resampling and coadding together FITS images.} to
align the combined NICMOS image to the ACS images.  The final SWarped
NICMOS F160W image with $0.03''$ pixel scales is shown in
Fig.~\ref{fig:B1608nicF160}.

\subsection{PSF modeling}
\label{sec:psf}
In this subsection, we describe the procedure for obtaining the PSFs for
each of the ACS and the NICMOS data sets.


\subsubsection{ACS PSF}

The ACS PSF is both spatially and temporally varying
\citep[e.g.][]{RhodesEtal07}.  One source of temporal variation is the
``breathing'' of the telescope while it orbits, which causes the focal
length (and, hence, the PSF) of the telescope to change.  Instead of
adopting a universal PSF, we take the approach of modeling several
PSFs using different means, and quantitatively comparing them using
the Bayesian analysis described in Section
\ref{sec:PPRMethod:matrix:probTheory}.  This has the advantage of
using the data (the observed image) to rank the models.  For each of
the two drizzled ACS images, we create five models for the PSF either based
on the TinyTim package \citep{KristHook97} or from the
unsaturated stars in the field: (1) drizzled PSF (``PSF-drz'') from a
set of TinyTim simulations \citep[following][]{RhodesEtal07}, (2)
single (nondrizzled) TinyTim PSF (``PSF-f3'') with a telescope focus
value of $-3$, (3) the closest star (``PSF-C'') located at $\sim
9''$ in the northeast direction from B1608+656 in the drizzled ACS
field with a Vega magnitude of 21.3 in F814W, (4) bright star \#1
(``PSF-B1'') that is located at $\sim 1.9'$ southwest of B1608+656 in
the drizzled ACS field with a Vega magnitude of 18.7 in F814W, and (5)
bright star \#2 (``PSF-B2'') that is located at $\sim 1.6'$ south of
B1608+656 in the drizzled ACS field with a Vega magnitude of 19.1.

The TinyTim frame(s) were drizzled and resampled to pixel sizes of
$0.03''$ to match the resolution of the ACS images.  We keep in mind
that the TinyTim PSFs (PSF-drz and PSF-f3) may be insufficient due to
the time-varying nature of the PSF and the aging of the detector
since the TinyTim code was written.  We expect the closest star to
B1608+656 (PSF-C) to be a good approximation to the PSF because the
spatial variation of the PSF across $\sim 9''$ should be negligible
and any temporal variations are the same as in the lens system.
However, this closest star is not bright enough to see the secondary
maxima in the PSF, so we additionally include two of the brightest
stars in the drizzled field mentioned above.  For each of the stars in
F606W and F814W, we make a small cutout around the star ($25\times25$
pixels for PSF-C, $51\times51$ pixels for PSF-B1, and $41\times41$
pixels for PSF-B2) and center it on a $200\times200$ grid, which is
the size of the drizzled science image cutouts of B1608+656 that are
used for the image processing.


\subsubsection{NICMOS PSF}

The NICMOS PSF is thought to be more stable, and thus we assume a
TinyTim model for it.  The output TinyTim PSF is in the CCD frame of
NICMOS with pixel size $0.043''$.  As with the F160W science image,
the PSF was SWarped to be aligned with the ACS images with $0.03''$
pixels.  Since there is only one PSF model for NICMOS, PSF
specifications throughout the rest of this paper refer to the ACS
PSFs.


\subsection{Dust correction}
\label{sec:dustCorr}

With observations in two or more wavelengths, we can correct for the
dust extinction using empirical dust extinction laws.  We adopt the
extinction law of \citet{CardelliEtal89} with the following dust
extinction ratios at the redshift of the lens $z_{\rm d}=0.63$ for $R_V=3.1$
(Galactic extinction): 
${A_{\rm{F606W}}}/{A_{\rm{V}}}=1.56$,
${A_{\rm{F814W}}}/{A_{\rm{V}}}=1.14$, and
${A_{\rm{F160W}}}/{A_{\rm{V}}}=0.41$,
where $A_{\rm \lambda}$ is the extinction (difference between the
observed and intrinsic magnitudes) at wavelength $\lambda$.  These
dust extinction ratios agree with the values from the extinction law
in \citet{Pei92} to within $1.5\%$.  In order to correct for the
extinction, we need to know the intrinsic colors of the objects
(details in Section \ref{sec:dustCorr:IntColor}).  For each color type
of object (the lens galaxies, the source galaxy, and the AGN of source
galaxy), we denote the intrinsic color by $Q_F=(m_{F,
  \rm{intrinsic}}-m_{1,\rm{intrinsic}})$ where $F=1,\ldots, N_{\rm{b}}$
is in sequence from the reddest to the bluest wavelengths (by
construction $Q_{1}=0$), and $N_{\rm{b}}$ is the number of wavelength
bands used for dust correction.  Combining the dust extinction ratios
and the definition of intrinsic colors, we can model the observed
magnitudes at each image pixel in each of the wavelength bands $F$ in
terms of $A_V$ and the intrinsic magnitude of the reddest wavelength
band $m_{1,\rm{intrinsic}}$ as
\be
\label{eq:mobsInAvm1}
m_F \equiv m_{F,\rm{observed}} =  m_{1,\rm{intrinsic}} + Q_F + A_V k_F + n_F,
\ee
where $k_F \equiv {A_{F}}/{A_V}$ are constants given by the extinction
law and $n_F$ is the noise in the data of wavelength band $F$.  We can
solve for $A_V$ and $m_{1,\rm{intrinsic}}$ at each image pixel by
minimizing the following $\chi^2_{\rm{dust}}$ for each pixel:
\be
\label{eq:dustChi2}
\chi^2_{\rm{dust}} = \sum_{F=1}^{N_{\rm{b}}} \left(m_F - m_{1,\rm{intrinsic}} - Q_F - A_V k_F \right)^2.
\ee
We have weighted the images of the different bands equally because the
uncertainty associated with $m_F$ is negligible compared to that of
$Q_F$, and the uncertainties in $Q_F$ are of comparable magnitudes for
the different bands $F$ relative to the reddest.  The solution that
minimizes $\chi^2_{\rm{dust}}$ is
\bea
\label{eq:AvSolnToChi2}
A_V =& \bigg[ &\frac{1}{N_{\rm{b}}}\left(\sum_F k_F \right) \left(\sum_F m_F \right) - \nonumber \\
& & - \frac{1}{N_{\rm{b}}}\left(\sum_F k_F \right) \left(\sum_F Q_F \right) - \nonumber \\
& & - \sum_F k_F m_F + \sum_F k_F Q_F \bigg] \bigg/ \nonumber \\
& \bigg[&  \frac{1}{N_{\rm{b}}} \left(\sum_F k_F\right)^2 - \sum_F k_F^2 \bigg],
\eea
and
\be
\label{eq:m1SolnToChi2}
m_{1,\rm{intrinsic}} = \frac{1}{N_{\rm{b}}}\left(\sum_F m_F - \sum_F
  Q_F - \sum_F A_V k_F \right), 
\ee 
where the sums over $F$ go from $1,\ldots, N_{\rm{b}}$.  We emphasize
that Equations (\ref{eq:AvSolnToChi2}) and (\ref{eq:m1SolnToChi2})
give the $A_V$ and $m_{1,\rm{intrinsic}}$ at each pixel.  Since
$A_V$ varies from pixel to pixel (depending on the amount of dust seen
in that pixel), the various $A_V$ values of all pixels provide a dust map.
Similarly, the $m_{1,\rm{intrinsic}}$ values of all pixels give the
dust-corrected image in the reddest wavelength band.  The resulting
values of $m_{1,\rm{intrinsic}}$ and the intrinsic colors yield the
intrinsic (dust-corrected) magnitudes in the other bands
$m_{F,\rm{intrinsic}}$ where $F=2,\ldots, N_{\rm{b}}$.  For any one
band $F$, we can then construct the diagonal
dust matrix $\dustSet$ in Equation (\ref{eq:dataVecComp}) whose
nonzero entries are $10^{-0.4 A_V k_F}$.


\subsubsection{Obtaining the intrinsic colors}
\label{sec:dustCorr:IntColor}
The dust correction method outlined above requires the intrinsic
colors to be determined from the color maps.  To construct the color
maps, we need to unify the different resolutions of the images in
different bands (due to the wavelength dependence of the PSF).  We do
so by deconvolving the F606W, F814W, and F160W images using their
corresponding PSFs, and reconvolving the images with the F814W PSF for
each set of the five ACS PSFs and the single NICMOS PSF described in
Section \ref{sec:psf}.  Reconvolved images are preferred to
deconvolved images, because the latter show small-scale features (of a
few pixels' size) that are artificial due to the amplification of the
noise during the deconvolution process.  We select the F814W PSF for
the reconvolution because F814W will be used for the lens potential
modeling, due to its high S/N compared with F160W and
its less severe dust extinction compared with F606W.  In working with
the reconvolved images, we assume that the dust varies on a
scale larger than the F814W PSF, which is true for the regions near
the Einstein ring.  For the deconvolution, we use IDL's
\texttt{max\_entropy} iterative routine that is based on the algorithm
by \citet{HollisEtal92}.  We were unable to deconvolve the ACS F814W
image using PSF-f3.  This suggests that PSF-f3 is a bad model, which
we have expected due to temporal variations in the PSF.  PSF-f3 is a
single-epoch PSF whereas the F814W image was drizzled from multiple
exposures.  We, therefore, discard this PSF model.

\begin{figure*}
\begin{center}
\includegraphics[width=48mm]{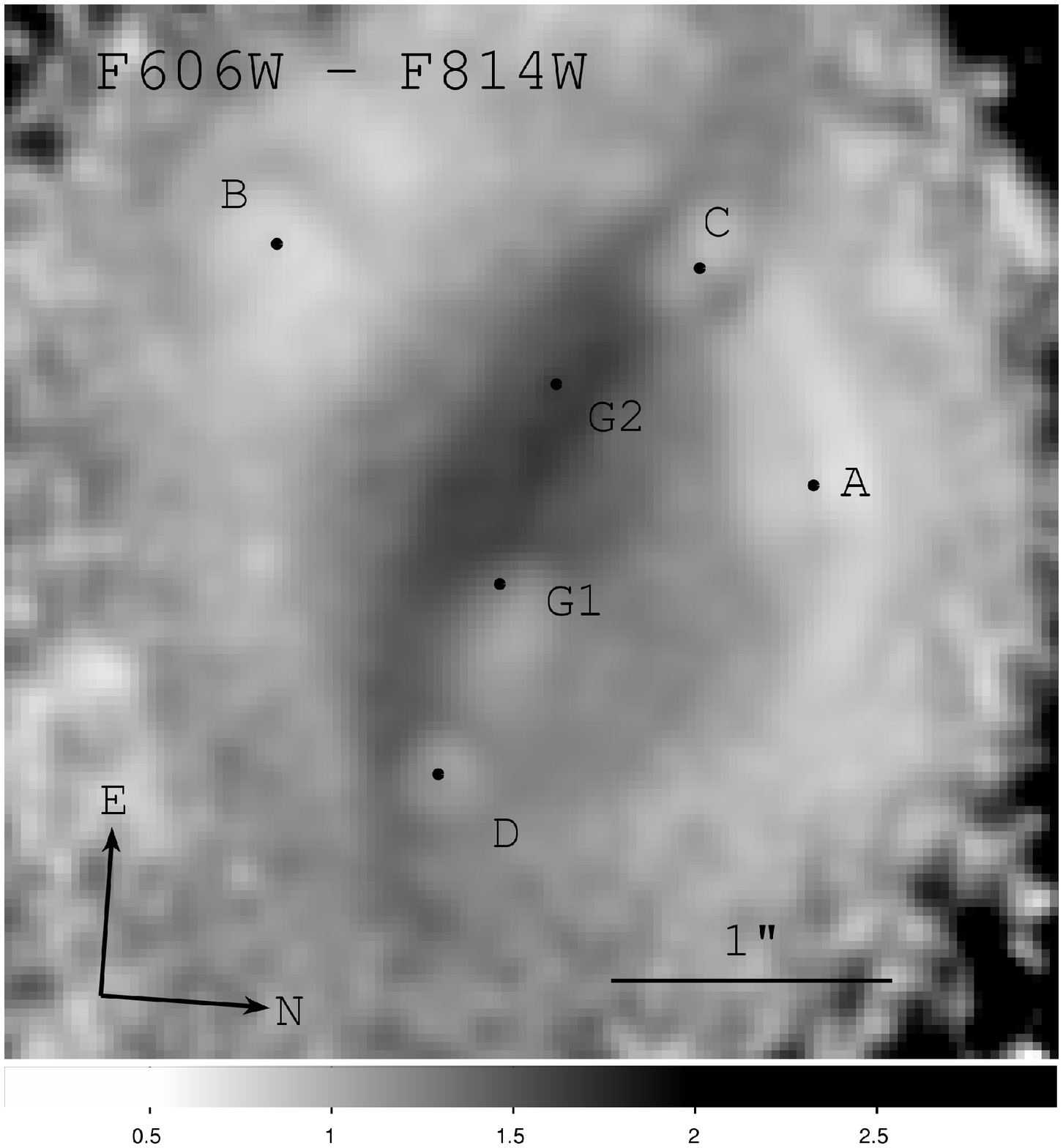}
\includegraphics[width=48mm]{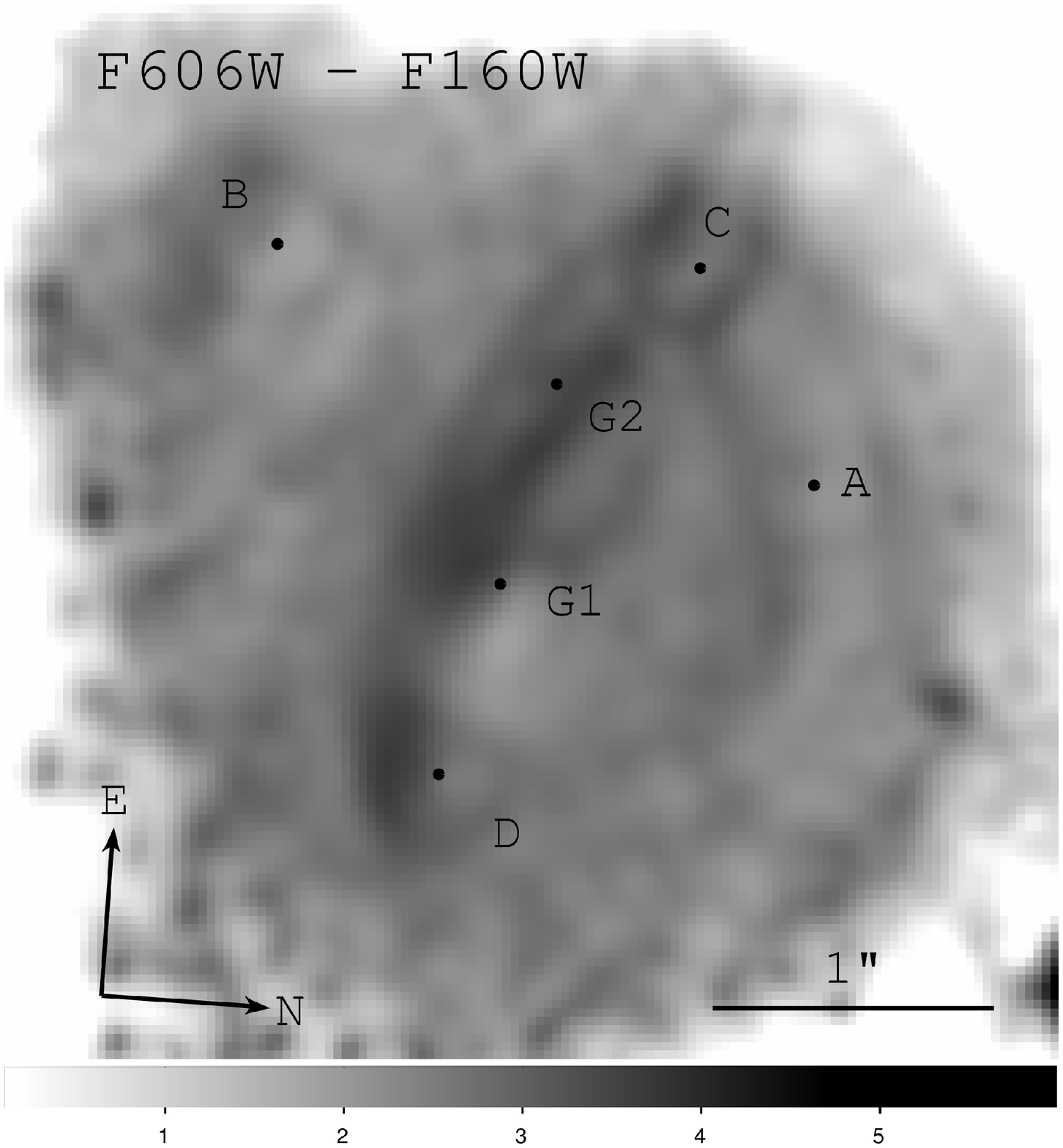}
\includegraphics[width=48mm]{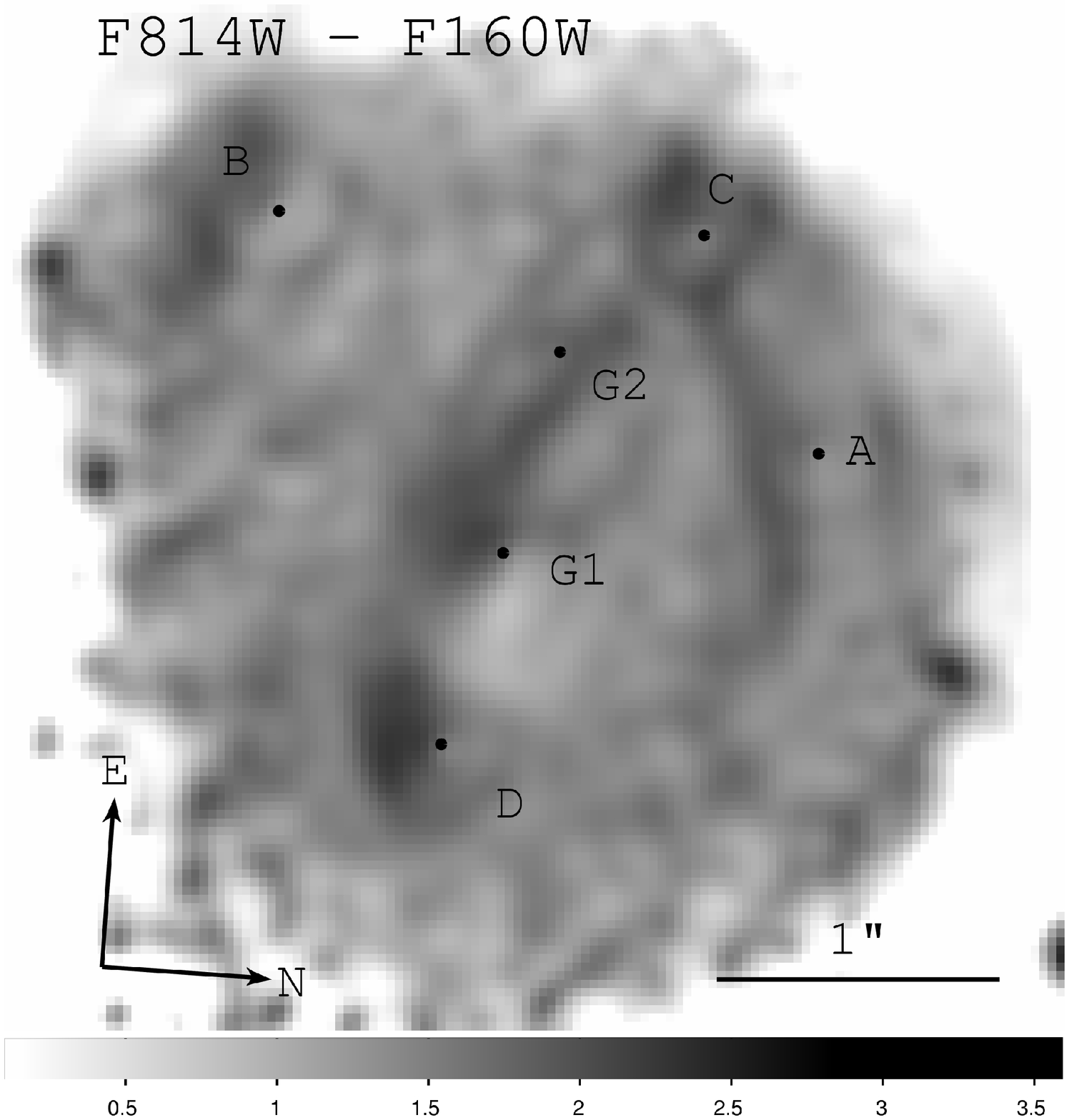}
\end{center}
\caption[Color maps from F606W, F814W, and F160W bands of B1608+656]{\label{fig:colorMapB1Star} From left to right: the derived color maps F606W--F814W, F606W--F160W, and F814W--F160 using PSF-B1.}
\end{figure*}

For each set of PSF models (PSF-drz, PSF-C, PSF-B1, and PSF-B2 for
ACS, and TinyTim PSF for NICMOS), we construct the color maps
F606W--F814W, F606W--F160W, and F814W--F160W from the reconvolved F606W,
F814W, and F160W images.  Fig.~\ref{fig:colorMapB1Star} shows the three
color maps derived for PSF-B1.  
Regions with bluer color slightly west of G1 are shown in all three color maps.  
Since the centroid of
this blue region is offset from the centroid of G1, we believe that
this blue region arises from differential reddening and not from
intrinsic color variations within G1, which is an elliptical galaxy
\citep{SurpiBlandford03}.  Since elliptical galaxies typically contain
little dust, \citet{KoopmansFassnacht99} and \citet{SurpiBlandford03}
suggested that the dust comes from G2, likely a dusty late-type
galaxy, through dynamical interaction.  This may explain why the
spectrum of G1 shows signatures of a young stellar population plus a
poststarburst population \citep{DresslerGunn83, MyersEtal95,
  SurpiBlandford03, KoopmansEtal03}: gas from G2 may have been
transferred to G1, where the tidal interactions may have triggered
star formation.

The color maps also show regions of bluer color around images C and D,
and we again believe that these are mostly differential reddening due to
the misalignment of the image positions and the centroids of these
blue regions, especially in F606W--F160W and F814W--F160W.  Furthermore,
we find more dust at the crossing point of the isophotal separatrix
(the figure-eight-shaped intensity contour) of the image pair
A--C. This is encouraging, as lensing models indeed predict the
crossing point to be closer to image A (see discussion in Section \ref{sec:resultDustMaps}).  
However, these bluer regions near
images C and D may also arise from the lensed source being
intrinsically bluer than the surrounding emission.  The F814W--F606W
color for these blue regions is consistent with typical star-forming
galaxies \citep[e.g.][]{ColemanEtal80}.  In the F606W--F814W color map,
there is a faint ridge of redder color connecting images A and C.
This may be due to the asymmetry in the stellar PSF model (with the
star position not exactly centered within a pixel), which would cause
the F606W and F814W isophotes to shift relative to each other after
the deconvolution and reconvolution.  For the color maps from the
other PSF models, we find that the color maps from PSF-C and PSF-B2
look similar to that from PSF-B1 with varying amounts of noise due to
varying brightnesses of the stellar PSFs.  PSF-drz gave color maps that
differ from those from the stellar PSFs (PSF-C, PSF-B1 and PSF-B2)
because PSF-drz, especially in the F606W band, did not exhibit a
single brightness peak but a string of equal brightness pixels at the
center due to frame alignment difficulties during the drizzling
process.  This caused the brightest pixels in the Einstein ring to
shift by $\sim 1$ pixel after the deconvolution and the reconvolution
process in F606W, and created artificial sharp highlights tracing the
edge of the ring in the F606W--F814W color map.  As will be seen in
Section \ref{sec:ImProcModelComp}, this leads to PSF-drz and its
resulting dust map giving a lower goodness of fit in the lens
inversion, and hence being ranked lower compared with other models.

In each of the color maps, we define three color regions for the three
color components: one within the Einstein ring for the lens galaxies
(we assume G1 and G2 to have the same colors), one for the Einstein
ring of the lensed extended source, and one for the lensed AGN (core
of the extended source).  Following \citet{KoopmansEtal03}, we
determine the bluest color within each region, assume that this part
of the region was not absorbed by dust, and adopt this color as the
intrinsic color.  This assumes that each of the three components has a
constant intrinsic color.  This would allow us to obtain the
\textit{differential} reddening for each of the components across the
lensed image; \textit{absolute} reddening is not needed because a
uniform dust screen does not affect lens modeling.  Table
\ref{tab:intrinsicColors} lists the intrinsic colors for each of the
three pairs of color maps.  The intrinsic colors of F606W--F814W are
not identical to the difference between F606W--F160W and F814W--F160W,
but agree within the uncertainties (0.02--0.1).

\begin{table}
\begin{center}
\caption[Intrinsic colors of the AGN, Einstein ring, and lens galaxies in B1608+656]{\label{tab:intrinsicColors} Intrinsic colors of the AGN, Einstein ring, and lens galaxies in B1608+656
}
\begin{tabular}{llccc}
\tableline
\tableline
  & & F606W--F814W & F814W--F160W & F606W--F160W \\
\tableline
PSF-drz & AGN & $0.50$ & $1.4$  & $1.91$ \\
  & Ring & $0.70$      & $1.5$  & $2.20$ \\
  & Lens & $0.84$      & $1.0$  & $1.88$ \\
PSF-C & AGN & $0.78$   & $1.3$  & $2.10$ \\
  & Ring & $0.84$      & $1.5$  & $2.30$ \\
  & Lens & $1.04$      & $1.0$  & $2.05$ \\
PSF-B1 & AGN & $0.72$  & $1.1$  & $1.85$ \\
  & Ring & $0.76$      & $1.3$  & $2.10$ \\
  & Lens & $1.04$      & $0.82$ & $1.85$ \\
PSF-B2 & AGN & $0.70$  & $1.17$ & $1.99$ \\
  & Ring & $0.80$      & $1.3$  & $2.10$ \\
  & Lens & $1.01$      & $0.85$ & $1.92$ \\
\tableline
\end{tabular}
\end{center}
\tablecomments{The intrinsic colors are based on color maps derived from the four ACS PSF models (PSF-drz (drizzled TinyTim), PSF-C (closest star), PSF-B1 (bright star \#1), and PSF-B2 (bright star \#2)) and the single NICMOS TinyTim PSF.  The intrinsic colors for each of the three color regions are determined from the bluest colors in the respective region.  The uncertainties on the intrinsic colors vary from 0.02 to 0.1.  The higher uncertainties are associated with the F160W image, which has a lower S/N.}
\end{table}


\subsubsection{Resulting dust maps}
\label{sec:resultDustMaps}
With the intrinsic colors determined for each PSF model, we obtain two
dust maps ($A_V$ maps) using (1) only the ACS F606W and F814W images
and (2) the ACS F606W and F814W images together with the NICMOS F160W
image.  In this way, we can assess whether the inclusion of the lower
S/N NICMOS image (with the much broader PSF) improves the
dust correction.  

\begin{figure*}
\begin{center}
\includegraphics[width=75mm]{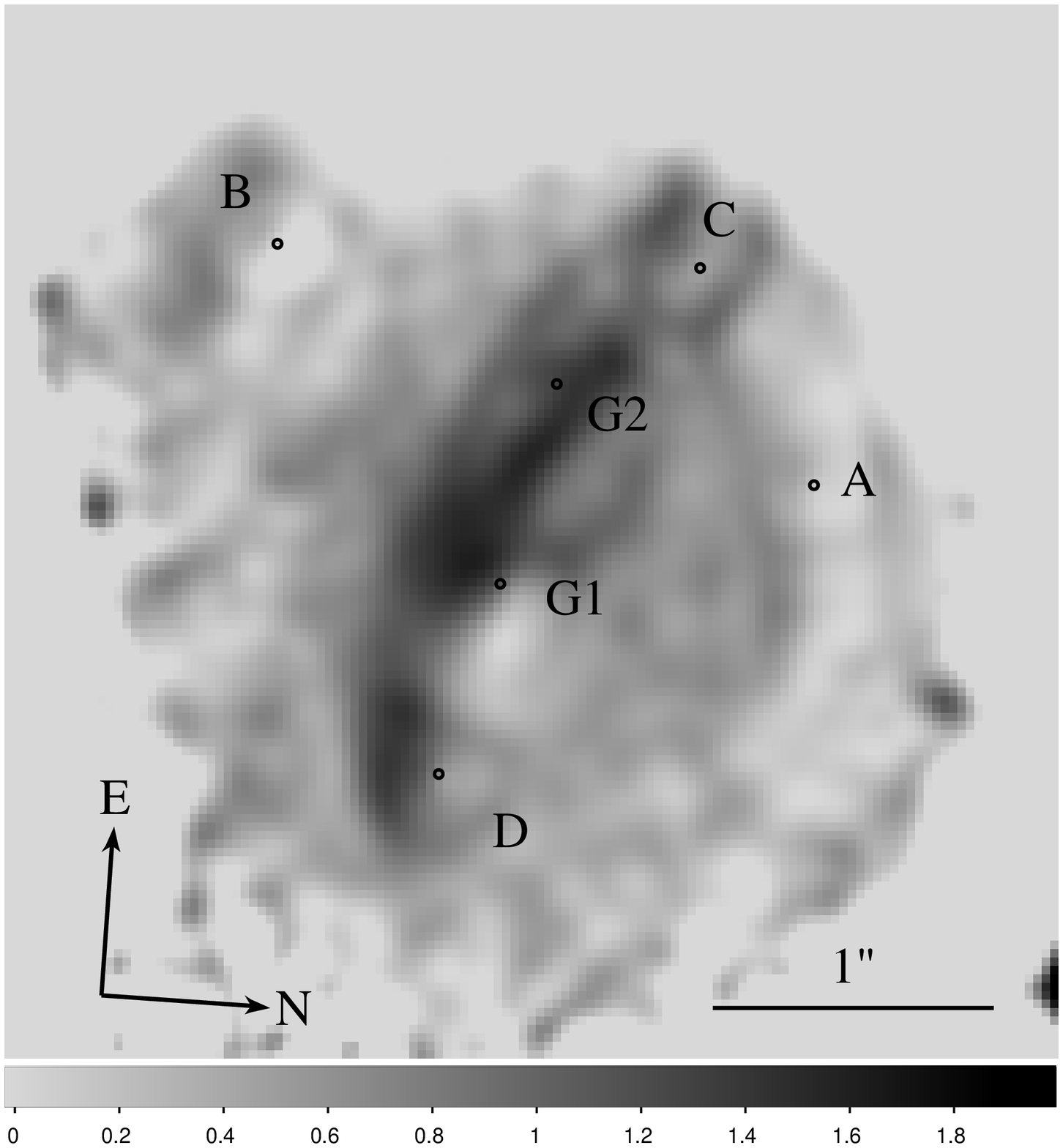}
\includegraphics[width=75mm]{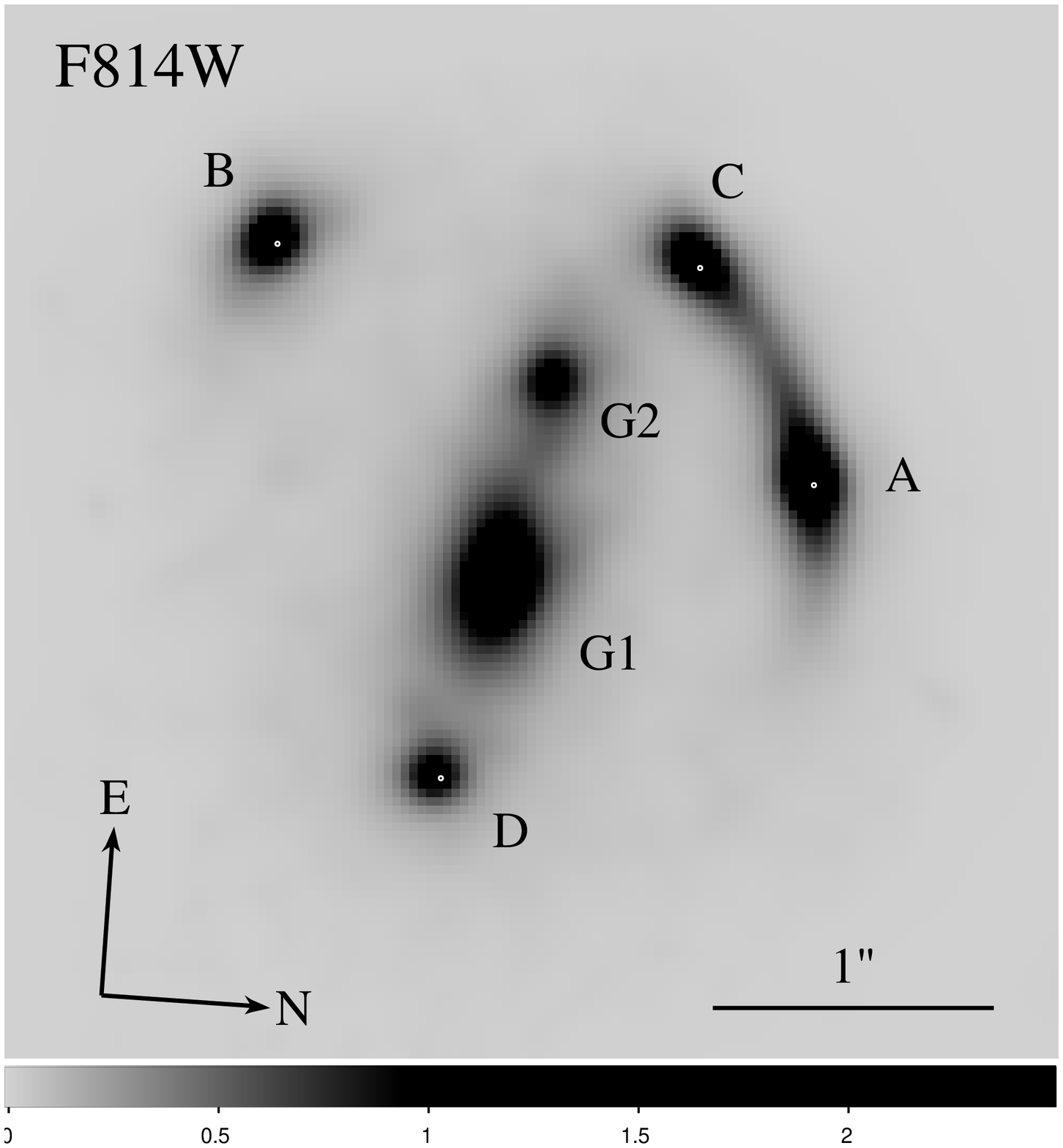}
\end{center}
\caption[Dust map and dust-corrected F814W image]{\label{fig:AVcorrF814WB1Star3band} Left-hand panel: the $A_V$ map obtained from dust correction with PSF-B1 using all three bands of images and the intrinsic colors listed in Table \ref{tab:intrinsicColors}.  The galactic dust extinction law was assumed.  The dust lane through images C, G2, G1, and D is visible.  Right-hand-panel: dust-extinction-corrected F814W image using PSF-B1 and the three-band dust map in the left-hand panel.  Compared to the right-hand panel in Fig.~\ref{fig:B1608acsF606F814}, the light profile of G1 is more elliptical and the crossing point of the isophotal separatrix of images A and C has shifted toward A after the dust correction.}
\end{figure*}

\begin{figure}
\begin{center}
\includegraphics[width=80mm]{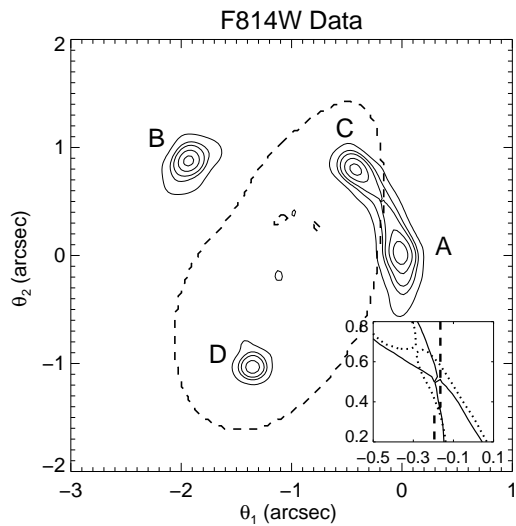}
\end{center}
\caption[Overlay of the dust-corrected and galaxy-subtracted image and the lens
potential]{\label{fig:overlayF814W} 
Crossing isophotes of the B1608+656 Einstein ring. Shown here is  the
dust-corrected and galaxy-subtracted F814W image (solid contours), with the
critical curves of the SPLE1+D (isotropic) potential model
\citep{KoopmansEtal03} overlaid (dashed curves).  The inset shows a
``zoomed-in'' view of the region between images C and A; here, the dotted curves
in the zoomed-in panel are the intensity contours of the  galaxy-subtracted
F814W image \emph{without} dust correction.  After dust correction, the 
crossing point of the isophotal separatrix (the center of the figure-eight
isophote)  is shifted toward the critical curve, indicating successful dust
correction.}

\end{figure}

The left-hand panel of Fig.~\ref{fig:AVcorrF814WB1Star3band} is the
resulting $A_V$ dust map derived using PSF-B1 and using images in all
three bands.  The dust map shows the east-west dust lane through the
system (absorbing light from C, G2, G1, and D) that is visible in the
original drizzled ACS F606W and F814W images.  There is little
extinction near images A and B, but there are faint rings surrounding
the images that are mostly due to imperfect F160W deconvolution.  We
note that the low S/N exterior to the Einstein ring
results in the dust map being noisy in this area.  We make sure that
these noisy areas are not included in the Bayesian evidence
computations in Sections \ref{sec:ImProcModelComp} and
\ref{sec:PotRec:B1608}.  The right-hand panel of
Fig.~\ref{fig:AVcorrF814WB1Star3band} is the resulting dust-corrected
F814W image that exhibits two signs of proper dust correction: the
correctly shifted crossing point of the isophotal separatrix of the
image pair A--C, as shown more clearly in Fig.~\ref{fig:overlayF814W},
and the smoother lens galaxy profiles.  As a result of recovering the
absorbed light, the dust-corrected image has higher intensity values
than the uncorrected image.  Therefore, we create a weight map for the
dust-corrected image by scaling the multidrizzle weight image in order to keep
the S/N of each pixel the same (before and after
dust correction).  This ``dust-corrected weight image'' will be used
in the next section for determining the lens galaxy light.

The dust maps obtained from the other PSF models with or without the
inclusion of the NICMOS image show similar features except for the
following two dust maps. 
\begin{enumerate}
\item The ACS-only (no NICMOS) dust map from PSF-B2 showed a
faint ridge of dust connecting images A and C.  As explained, this may
be due to the asymmetrical/bad PSF model.  Since the dust map otherwise
exhibits the correct features, we keep this dust map for the next
analysis step.
\item The ACS-only dust map from PSF-drz
showed prominent artificial lensing arc features due to the
$\sim 1$ pixel offset in the image positions/arcs in the deconvolved
and reconvolved F606W and F814W images, respectively.  Therefore, we discard this
dust map of the ACS-only images for PSF-drz, but keep the dust map
derived from using all three bands (that includes NICMOS).
\end{enumerate}

After discarding the ACS PSF-f3
and the ACS-only dust map from PSF-drz, we have a total of
seven dust maps (and resulting dust-corrected F814W images).  
All of these are reasonable
dust corrections to use since they are derived using representative PSFs and intrinsic colors.  We will compare these dust maps and PSF
models in Section \ref{sec:ImProcModelComp}.

\subsection{Lens galaxy light}
\label{sec:lensGalLight}

\begin{figure*}
\begin{center}
\includegraphics[width=48mm]{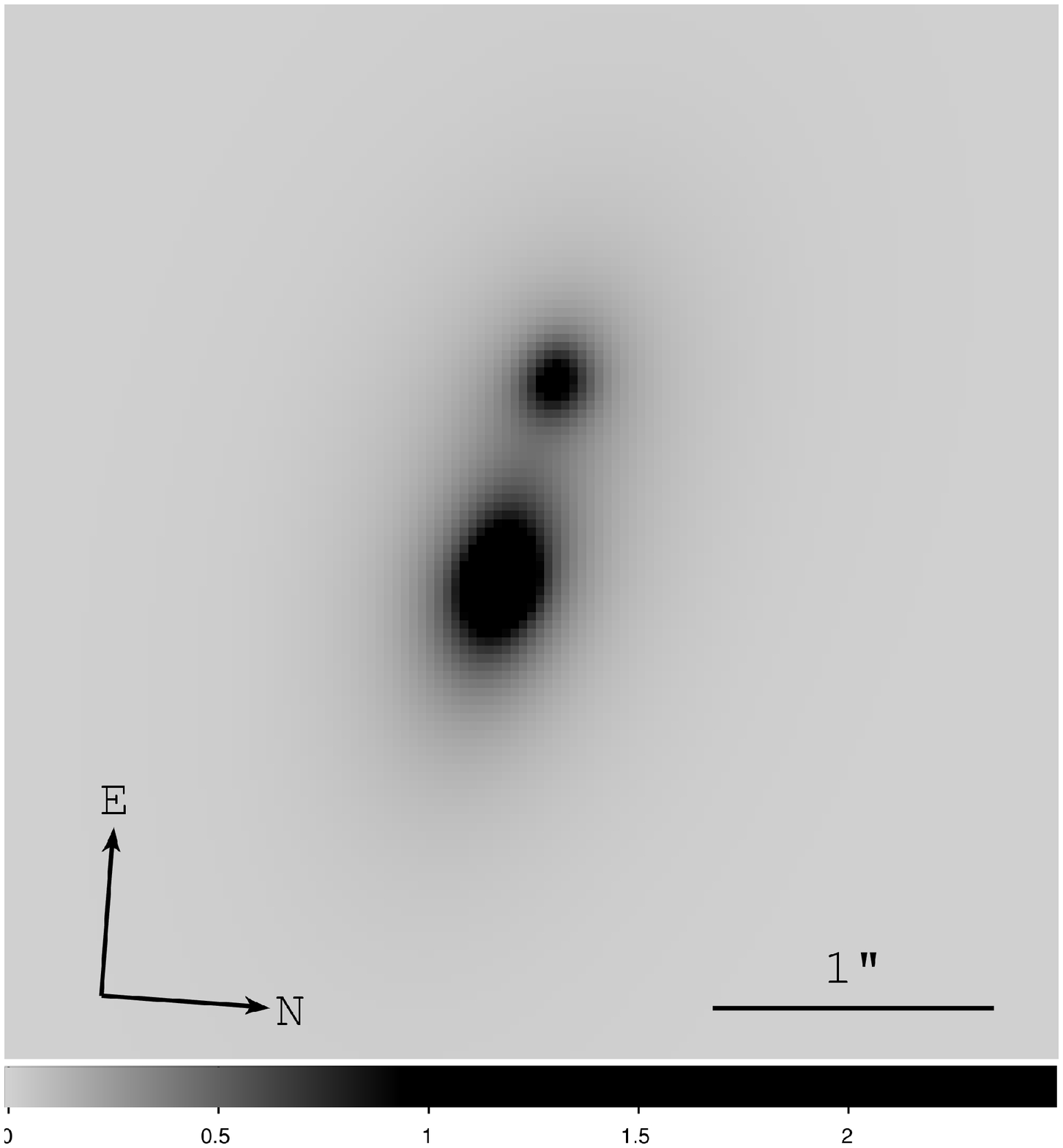}
\includegraphics[width=48mm]{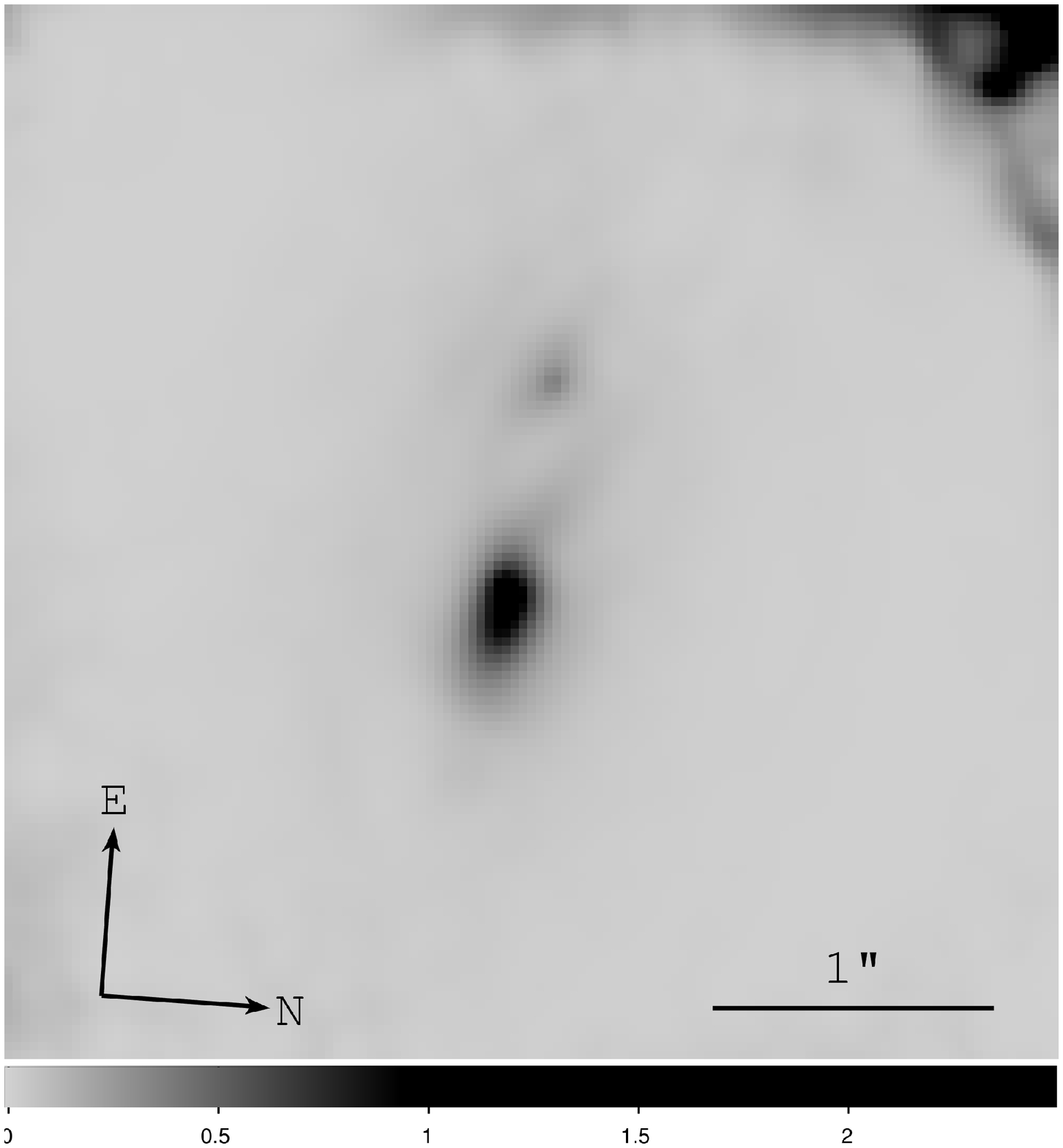}
\includegraphics[width=48mm]{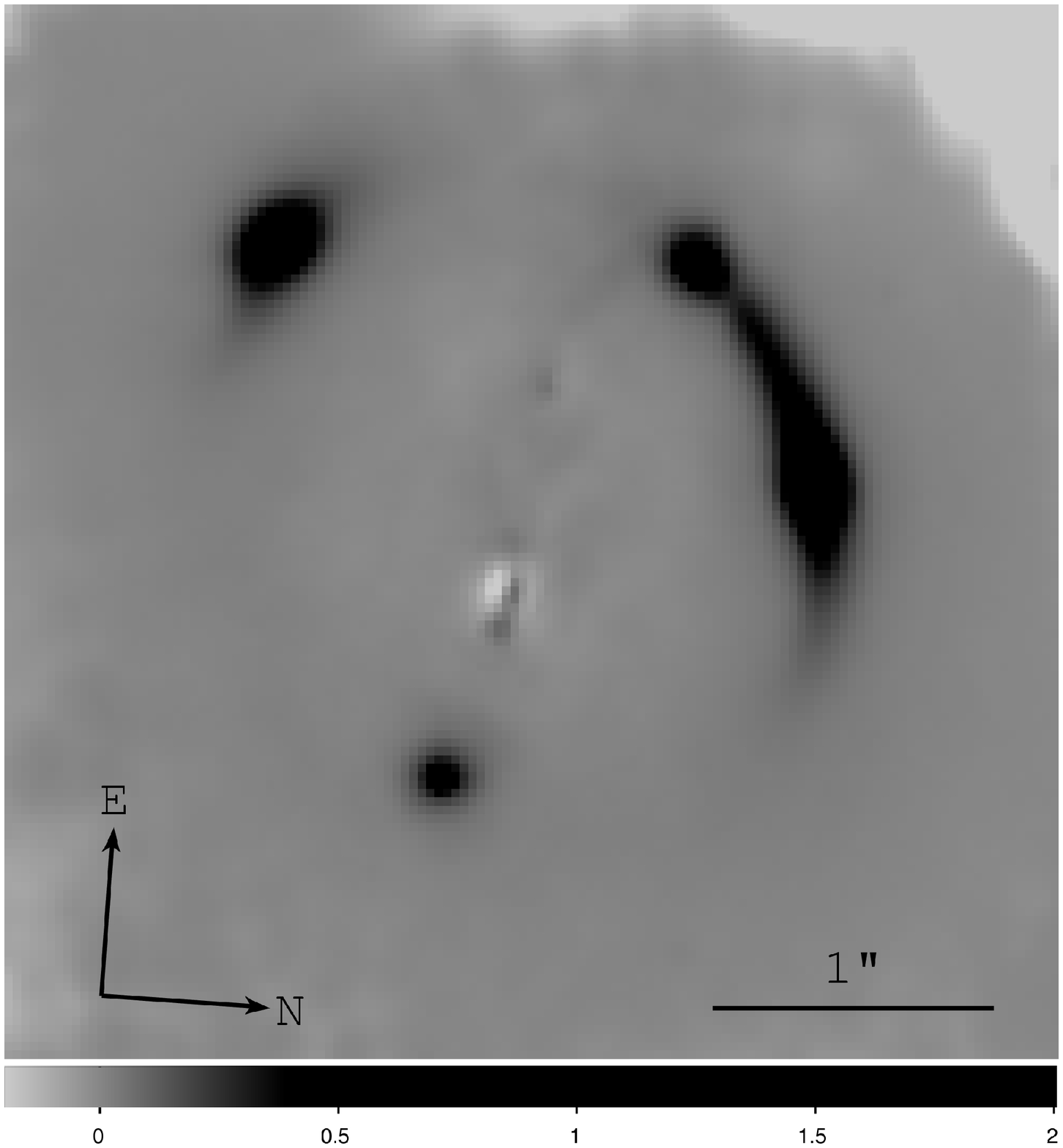}
\end{center}
\caption[Lens galaxy light fitting using S\'ersic profiles]
{\label{fig:galfitB1Star3band} S\'ersic lens galaxy light profile
  fitting to the dust-corrected F814W image, with PSF-B1 and its
  corresponding three-band dust map, using GALFIT.  The left-hand panel
  shows the best-fit S\'ersic light profiles with S\'ersic indices
  $(n_{\rm{G1}},n_{\rm{G2}})=(3,4)$.  The middle panel shows the dust-extincted 
  galaxy light profiles, which is the left-hand panel with
  the dust extinction added back in.  The right-hand panel shows image
  residual (difference between the F814W drizzled image in
  Fig.~\ref{fig:B1608acsF606F814} and the middle panel) with misfit
  near the cores of the lens galaxies of $\sim 25-35\%$.}

\end{figure*}

For each of the seven resulting dust-corrected F814W images in Section
\ref{sec:resultDustMaps} and its corresponding PSF, we create an
elliptical mask for the lens galaxies' region that excludes the
Einstein ring, and fit the lens galaxies' light to elliptical S\'ersic
profiles using GALFIT \citep{PengEtal02}.  In particular, we impose
the S\'ersic indices to be one of the following pairs: $(n_{\rm{G1}},
n_{\rm{G2}}) = (1,1), (2,2), (3,3), (3,4), (4,3), (4,4)$.  There are
more pairings with $n=3$ and $n=4$ since previous works by, for
examples, \citet{BlandfordEtal01} and \citet{KoopmansEtal03} found G1
to be well described by $n=4$ (de Vaucouleurs profile).  With the
dust-corrected weight image, we obtain a reduced $\chi^2$ value for
each of the profile fittings.  For each dust-corrected F814W image, we
pick the S\'ersic index pair with the lowest reduced $\chi^2$ from the
fit (top two pairs in the case of PSF-drz) and list it in Table
\ref{tab:galfitChi2}.  As an illustration,
Fig.~\ref{fig:galfitB1Star3band} shows the GALFIT S\'ersic
$(n_{\rm{G1}}, n_{\rm{G2}}) = (3,4)$ results of the dust-corrected
F814W image using the three-band dust map from PSF-B1.  The dark (light)
patches in the upper right-hand corner of the middle (right-hand)
panel result from the noisy dust map due to low signal to noise in
this area.  Apart from this area and the lens galaxies' cores, most of
the observed lens galaxies' light matches the dusted S\'ersic profiles
in the middle panel, as shown in the residual map in the right-hand
panel.  The misfit near the cores could be due to intrinsic color
variations in the lens galaxies, the dust screen assumption, PSF
imperfections, and/or inapplicability of a single S\'ersic model at
the center.  Nonetheless, accurate light fitting near the cores of the
lens galaxies is not important; it is for the isophotes of the
Einstein ring that we need to have accurate dust and lenses' light
corrections for the lens modeling.  For the ring, the dust screen
assumption in our approach is valid.

\begin{table}
\begin{center}
\caption[Best-fit S\'ersic light profiles for the lens galaxies]{\label{tab:galfitChi2} Best-fitting S\'ersic light profiles for the lens galaxies G1 and G2 for the seven different dust-corrected F814W images based on different PSF and dust maps
}
\begin{tabular}{ccccc}
\tableline
PSF & Dust Map & S\'ersic Indices $(n_{\rm{G1}}, n_{\rm{G2}})$ & Reduced $\chi^2_{\rm{lens\ light}}$ \\
\tableline
drz & Three-band & $(3,4)$ & 4.48 \\
drz & Three-band & $(3,3)$ & 4.53 \\
C & Three-band & $(3,4)$ & 5.11 \\
C & Two-band & $(3,3)$ & 6.13 \\
B1 & Three-band & $(3,4)$ & 5.53 \\
B1 & Two-band & $(2,2)$ & 7.16 \\
B2 & Three-band & $(2,2)$ & 5.95\\
B2 & Two-band & $(2,2)$ & 8.19\\
\tableline
\end{tabular}
\end{center}
\tablecomments{In the PSF column, ``drz'' = drizzled TinyTim, ``C'' = closest star, ``B1'' = bright star \#1, and ``B2'' = bright star \#2.  In the dust map column, ``two-band'' represents the dust map obtained from just the two ACS bands, and ``three-band'' represents the dust map obtained from the two ACS and the one NICMOS band.}
\end{table}

\subsection{Comparison of PSF, dust, and lens galaxy light models}
\label{sec:ImProcModelComp}

Following the method outlined in Section \ref{sec:PPRMethod:realData},
we can use the Bayesian evidence from the source-intensity
reconstruction to compare the different PSF ($\blurSet$), dust
($\dustSet$) and lens galaxy light ($\glightVec$) models.  For each
set of $\blurSet$, $\dustSet$, and $\glightVec$, we obtain the
corresponding galaxy-subtracted F814W image
($\dataVec-\blurSet\cdot\dustSet\cdot\glightVec$) that is analogous to
the one shown in the right-hand panel of
Fig.~\ref{fig:galfitB1Star3band}.  We then make a $130\times130$ pixel
cutout of the $0.03''$ galaxy-subtracted image and use the SPLE1+D
(isotropic) lens potential model in \citet{KoopmansEtal03}, which is
the most up-to-date simply-parameterized lens potential model for
B1608+656, for the source-intensity reconstruction.  Due to the source
and image pixelizations, we include the regridding error (described in
Section \ref{sec:ACSimprocess}) in the image covariance matrix.

We select an annular region enclosing the Einstein ring, and use the
data inside this region for the source-intensity reconstructions for
each set of the PSF, dust, and lens galaxy light models.  The source
grid, which we fix to have $32\times32$ pixels, has pixel sizes that
are $\sim 0.022''$ to cover the marked elliptical annular region when
mapped to the image plane.  This is sufficient for achieving
reasonable reconstructions and is computationally manageable.  In the
inversions, we reduced the PSF to $15\times15$ pixels to keep the
matrices such as $\blurSet$ reasonably sparse for computing speed.  We
try three forms of regularization: zeroth-order, gradient and
curvature (e.g. Appendix A of \citeauthor{SuyuEtal06}
\citeyear{SuyuEtal06}).

\begin{figure*}
\begin{center}
\includegraphics[width=150mm]{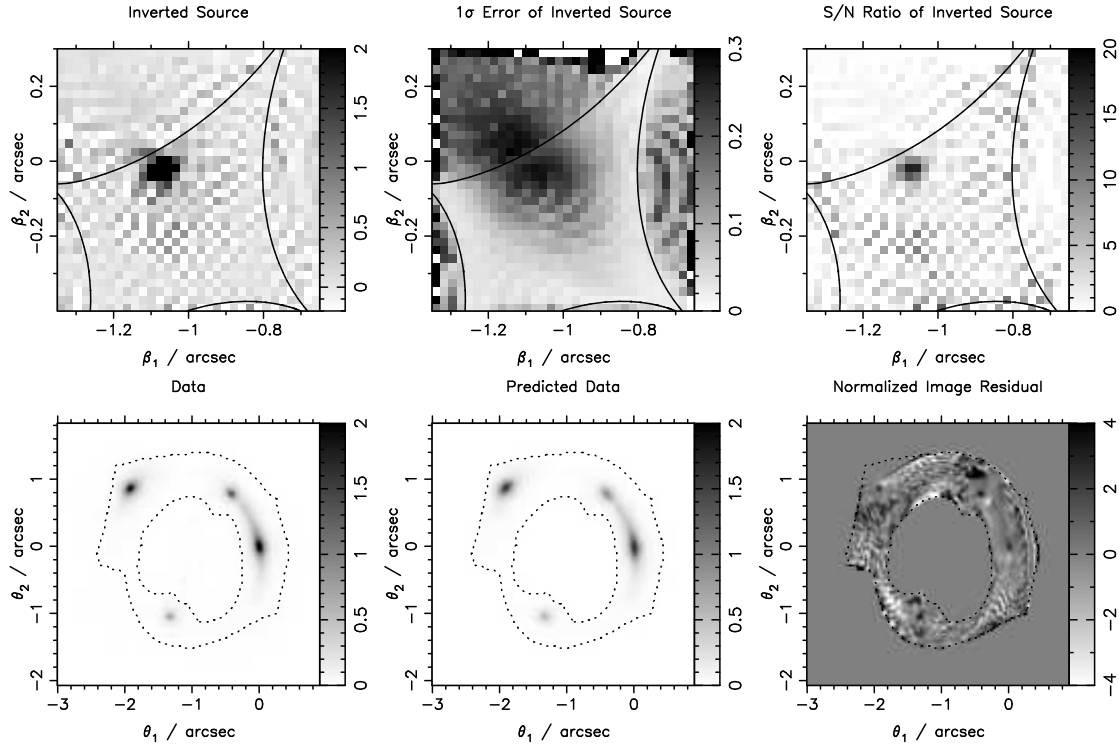}
\end{center}

\caption[source-intensity reconstruction of B1608+656 with PSF-B1, its corresponding three-band dust map and lens galaxy light]
{\label{fig:B1608SrRecB1Star3band} Source-intensity reconstruction of
B1608+656  (assuming model \#5 in Table \ref{tab:B1608SingleSrRecEvid}). 
Top panels from left
to right: the reconstructed source-intensity distribution with the caustic
curves of the SPLE1+D (isotropic) model overlaid, 
the $1\sigma$ error for the source-intensity values, 
the S/N of the reconstruction (i.e., 
the ratio of the top
left-hand to the top-middle panel).  
Bottom panels from left to right: the observed
F814W galaxy-subtracted image, the reconstructed image using the reconstructed
source in the top left-hand panel, and the normalized image residual (i.e., 
the map of the 
difference between the bottom left-hand and the bottom middle panels, in units
of the estimated pixel uncertainty from the data image covariance matrix).}

\end{figure*}

Table \ref{tab:B1608SingleSrRecEvid} lists the suite of PSF, dust, and
lens galaxy light models we obtained in the previous section.  We
label the different models by numbers from 1 to 11 in the left-most
column.  Models 9 and 10 correspond to the mixing of the dust maps and
lens galaxy light profiles derived from PSF-B1 with PSF-C and vice
versa.  Model 11, which is included as a consistency check, uses
PSF-B1 and has no dust correction applied.  For each set of models,
the source-intensity distribution for B1608+656 is reconstructed.  As
an example, Fig.~\ref{fig:B1608SrRecB1Star3band} shows the results of
the source reconstruction with gradient regularization using PSF-B1,
its corresponding three-band dust map, and the resulting S\'ersic
($n_{\rm{G1}},n_{\rm{G2}})=(3,4)$ galaxy light profile.  The top
left-hand panel shows the reconstructed source-intensity distribution
that is approximately localized, an indication that the lens potential
model is close to the true potential model.  In the top-middle panel,
the pixels that are far from the source but are inside the
caustics have lower $1\sigma$ error values than the pixels outside the
caustics due to higher image multiplicity inside the caustics.  The
bottom right-hand panel shows significant image residuals (the reduced
$\chi^2$ is 1.9 inside the annulus), a sign that the PSF, dust, lens
galaxy light, and/or the lens potential models are not optimal.  In
Section \ref{sec:PotRec:B1608}, we will use the pixelated potential
correction scheme, which is more suitable for interacting galaxy
lenses, to improve the simply-parameterized SPLE1+D (isotropic) model.

The source-intensity reconstructions using other PSF and lens galaxy
light models with three-band dust maps give overall similar inverted
source intensities and image residuals, but the source intensities can
be more or less localized and the magnitude and structures of the
image residuals vary for different model sets.  However, the
source-intensity reconstructions using models with two-band dust maps
result in source intensities that are not localized, and the image
residuals show surpluses of light in the ring region and deficits of
light in the lens galaxy region (corresponding to the color regions we
marked for obtaining the intrinsic colors).  The reason is that with
only two bands, the resulting dust-corrected F814W image is highly
sensitive to relative shifts between the F606W and F814W images (due to an
imperfect PSF model, deconvolution, and reconvolution) and errors in
the modeled intrinsic colors.  The abrupt change in the modeled
intrinsic colors across the boundaries of the color regions creates
artificial surpluses or deficits of dust-corrected light near the
boundaries.  This effect is suppressed with the addition of the F160W
image because the F160W image suffers relatively little extinction,
and the error due to misalignment in the images and abrupt change in
the modeled intrinsic colors is reduced when one has more than two
bands.  A few tests suggest that the error in the dust-corrected image
due to the range of intrinsic colors listed in Table
\ref{tab:intrinsicColors} overwhelms the error associated with the
foreground dust screen assumption for the lens galaxy light.

The source-intensity reconstruction in Model 11 with no dust
correction shows significant image residuals in the extended ring,
with overall surpluses of light surrounding images A and B and deficits
surrounding images C and D.  The source intensity is also poorly
reconstructed, being nonlocalized and noisy.  This illustrates the
importance of dust correction for the initial SPLE1+D (isotropic)
model.

\begin{table}
\begin{center}
\caption[PSF, dust, and lens galaxies' light model comparison based on Bayesian source inversion]{\label{tab:B1608SingleSrRecEvid} PSF, dust, and lens galaxies' light model comparison based on Bayesian source inversion 
}
\begin{tabular}{cccccc}
\tableline
\tableline
 & PSF & Dust Map & S\'ersic $(n_{\rm{G1}}, n_{\rm{G2}})$ & Reg. Type & Log Evidence \\
 & & & & & ($\times 10^4$)\\
\tableline
1 & drz & Three-band & $(3,4)$ & grad & $1.49$ \\
2 & drz & Three-band & $(3,3)$ & grad & $1.48$ \\
3 & C & Three-band & $(3,4)$ & grad & $1.60$ \\
4 & C & Two-band & $(3,3)$ & zeroth & $1.40$ \\
5 & B1 & Three-band & $(3,4)$ & grad & $1.56$ \\
6 & B1 & Two-band & $(2,2)$ & zeroth & $1.10$ \\
7 & B2 & Three-band & $(2,2)$ & grad & $1.55$ \\
8 & B2 & Two-band & $(2,2)$ & zeroth & $1.23$ \\
9 & C & B1/three-band & $(3,4)$ & zeroth & $1.56$ \\
10 & B1 & C/two-band & $(3,3)$ & zeroth & $1.36$ \\
11 & B1 & --- & $(3,4)$ & zeroth & $1.27$ \\
\tableline
\end{tabular}
\end{center}
\tablecomments{For each set of the PSF, dust, and lens galaxy light profiles derived in Sections \ref{sec:psf}--\ref{sec:lensGalLight}, the Bayesian log evidence value is from the source-intensity reconstruction using the SPLE1+D (isotropic) model in \citet{KoopmansEtal03}.  The uncertainty in the log evidence value due to source pixelization is $\sim 0.03\times10^4$.  In the PSF column, ``drz'' = drizzled TinyTim, ``C'' = closest star, ``B1'' = bright star \#1, and ``B2'' = bright star \#2.  In the dust map column, we list ``two-band'' for the dust map obtained from just the two ACS bands and ``three-band'' for the dust map obtained from the two ACS and the one NICMOS band.  Unless otherwise indicated in the dust map column, the PSF model used for the dust map derivation was the same as the corresponding PSF model in the PSF column that was used for source reconstruction.  For completeness, we restate the S\'ersic indices in Table \ref{tab:galfitChi2} in the lens galaxy light profile column, which were obtained for the corresponding dust maps and PSFs specified in the dust map column.  The column of ``Reg.~Type'' refers to the preferred type of regularization for the source reconstruction, based on the highest Bayesian evidence value.  It can be one of three types: zeroth-order, gradient, or curvature.}
\end{table}

\subsubsection{Results of Comparison}
\label{sec:ImProcModelComp:CompResults}

Table \ref{tab:B1608SingleSrRecEvid} summarizes the results of model
comparison.  The ``Reg.~Type'' column denotes the preferred type of
regularization for the source reconstruction based on the highest
Bayesian evidence value \citep{SuyuEtal06}.  It can be one of the
three types that we use: zeroth-order, gradient, and curvature.  The
last column lists the log evidence values from the inversions.
Assuming the different models to be equally probable a
  priori, we use these evidence values for model comparison.  The log
evidence values range from $1.1\times10^4$ to $1.6\times 10^4$ with
uncertainties of $\sim 0.03\times10^4$ due to the finite source
resolution.

The list shows that the three-band dust models have higher evidence values
than the two-band dust models.  This is attributed to the two-band dust
models showing image residuals from the aforementioned artificial
surpluses and deficits of light in the dust-corrected image.  The
inclusion of the NICMOS F160W image to the ACS images (F606W and
F814W) for the dust correction is, therefore, crucial due to (1) the
proximity in the wavelengths of the ACS images and (2) the reduction
in the error associated with image misalignments and simplistic
intrinsic color models.

The three-band dust models also have higher evidence values than the
no-dust model.  This further validates the three-band dust correction, as
already indicated by Fig.~\ref{fig:overlayF814W}.  The evidence value
of the no-dust model is in midst of the values for the two-band dust
models, suggesting that the systematic effects in the two-band dust maps
are comparable to the corrections that the dust maps are meant to
achieve, thus leading to little improvement in the lens modeling.

The difference between the evidence values in Models 1 and 2 (where
the models only differ in the S\'ersic light profiles) is, in general,
smaller than the difference between one of these two models and
another PSF/dust model.  Therefore, the source reconstruction (part of
lens modeling) seems to be less sensitive to the galaxy light profiles
than the PSF/dust models.  This is in agreement with our finding that
the dust-corrected image depends more on the PSF and the intrinsic
color models than on the form of the lens galaxy light and the dust
associated with the lens galaxies.  Models 1 and 2 with PSF-drz have
log evidence values on the low side of the collection of models with
three-band dust maps, which was expected with PSF-drz not having a single
brightness central peak due to misalignments in the drizzling process.
The other models with three-band dust maps (Models 3, 5, 7 and 9) have
effectively the same evidence values within the uncertainties.  The
models with two-band dust maps (Models 4, 6, 8, and 10) lead to a range
of evidence values with the PSF-C dust map being preferred to the
PSF-B1 and PSF-B2 dust maps.  The two-band dust maps suggest that the
shape of the primary maximum in the PSF is more important in the
modeling than the inclusion of secondary maxima since PSF-C, which we
expect to have a more accurate shape for the primary PSF maximum than
PSF-B1 and PSF-B2, does not have the secondary maxima whereas PSF-B1
and PSF-B2 do.  The asymmetry in the PSF due to the star not being
centered on a single pixel may also explain the less-preferred PSF-B1
and PSF-B2.  The distinction between the various stellar PSFs vanishes
with the three-band dust maps, possibly due to the higher amount of noise
in the three-band dust map with the inclusion of the lower S/N
NICMOS image.  In this case, the effects of the PSF variations across
the field are suppressed.

All models preferred either the zeroth-order or gradient form of
regularization, but never the curvature form; however, we mention that
the difference in the log evidence values between the different
regularization schemes ($\lesssim 3\times 10^2$) are on the order of
the uncertainties due to source pixelization, and the resulting
reconstructions for different types of regularizations are almost
identical.  This is because differences in evidence values between
models are currently dominated by changes in goodness of fit rather
than subtle differences between the prior forms.  Only when the image
residual is reduced will the prior (regularization) begin to play a
greater role in avoiding the reconstruction to fit to noise in the
data by keeping the source model simple.

This section has illustrated a method of creating sensible PSF, dust,
and lens galaxy light models for the gravitational lens B1608+656.  We
have obtained a representative sample of models, and have compared
these models quantitatively.  This collection of PSF, dust, and lens
galaxy light models leads to image residuals that cannot be beaten
down further unless we improve the SPLE1+D (isotropic)
simply-parameterized lens potential model by \citet{KoopmansEtal03} to
take into account the two \textit{interacting} galaxy lenses.  The
pixelated potential reconstruction of B1608+656 is the subject of Section
\ref{sec:PotRec:B1608}.


\section{Pixelated lens potential of B1608+656}
\label{sec:PotRec:B1608}

We reconstruct the lens potential for each set of the PSF, dust, and
lens galaxies' light in Models 2--11 in Table
\ref{tab:B1608SingleSrRecEvid}.  We describe in detail the potential
reconstruction using Model 5, which is one of the four models that,
within the uncertainties, have the highest Bayesian evidence value
before the potential correction.  At the end of the section, we
discuss the differences in the potential reconstruction between the
various PSF, dust, and lens galaxies' light models.

To reconstruct the lens potential of B1608+656, we use a
$130\times130$ pixel cutout of the drizzled ACS/F814W image with the pixel
size $0.03''$ shown in Fig.~\ref{fig:B1608acsF606F814}.  The
galaxy-subtracted F814W image
($=\dataVec-\blurSet\cdot\dustSet\cdot\glightVec$) is a $130\times130$
subimage of the right-hand panel in Fig.~\ref{fig:galfitB1Star3band}
with $200\times200$ pixels.

We follow the potential reconstruction method that was shown to
succeed in Section \ref{sec:PPRMethod:demo}.  For the initial lens
potential model, we use the SPLE1+D (isotropic) model from
\citet{KoopmansEtal03}. We perform nine iterations (labeled as 0--8) of
pixelated potential corrections on B1608+656.  For each iteration, we
first reconstruct the source intensity on a $32\times32$ grid with
pixel sizes of $0.022''$.  The source region is chosen so that it maps
to a completely joined annulus on the image plane (so that we can
determine the relative potential difference between images).  As in
Section \ref{sec:ImProcModelComp}, the PSF is reduced to a
$15\times15$ matrix to keep the inversion matrices sparse (and
computation time low).  Furthermore, we use only the curvature type of
regularization for the source reconstruction to reduce computation
time and to have regularized source-intensity gradients for the
potential corrections.  The source inversions are over-regularized in
the early iterations to ensure a smooth resulting source for taking
gradients.  The source over-regularization factors start at 1000 and
are gradually decreased to 1 at iteration=8.  With the resulting
source-intensity gradients and intensity deficits from the source
reconstruction, we perform the potential correction on a grid of
$30\times30$ pixels.  We use the curvature form of regularization for
each potential correction iteration.  To keep the corrections linear,
the potential corrections are also over-regularized with the
regularization constant ($\mu$) set at $10$ times the value where $\mu
E_{\rm{\dpsi}}$ peaks, as in Section \ref{sec:PPRMethod:demo}.  The
corrected potential has the midpoints in the left, bottom, and right
parts of the annular reconstruction region fixed to the initial
potential model.

\begin{figure*}
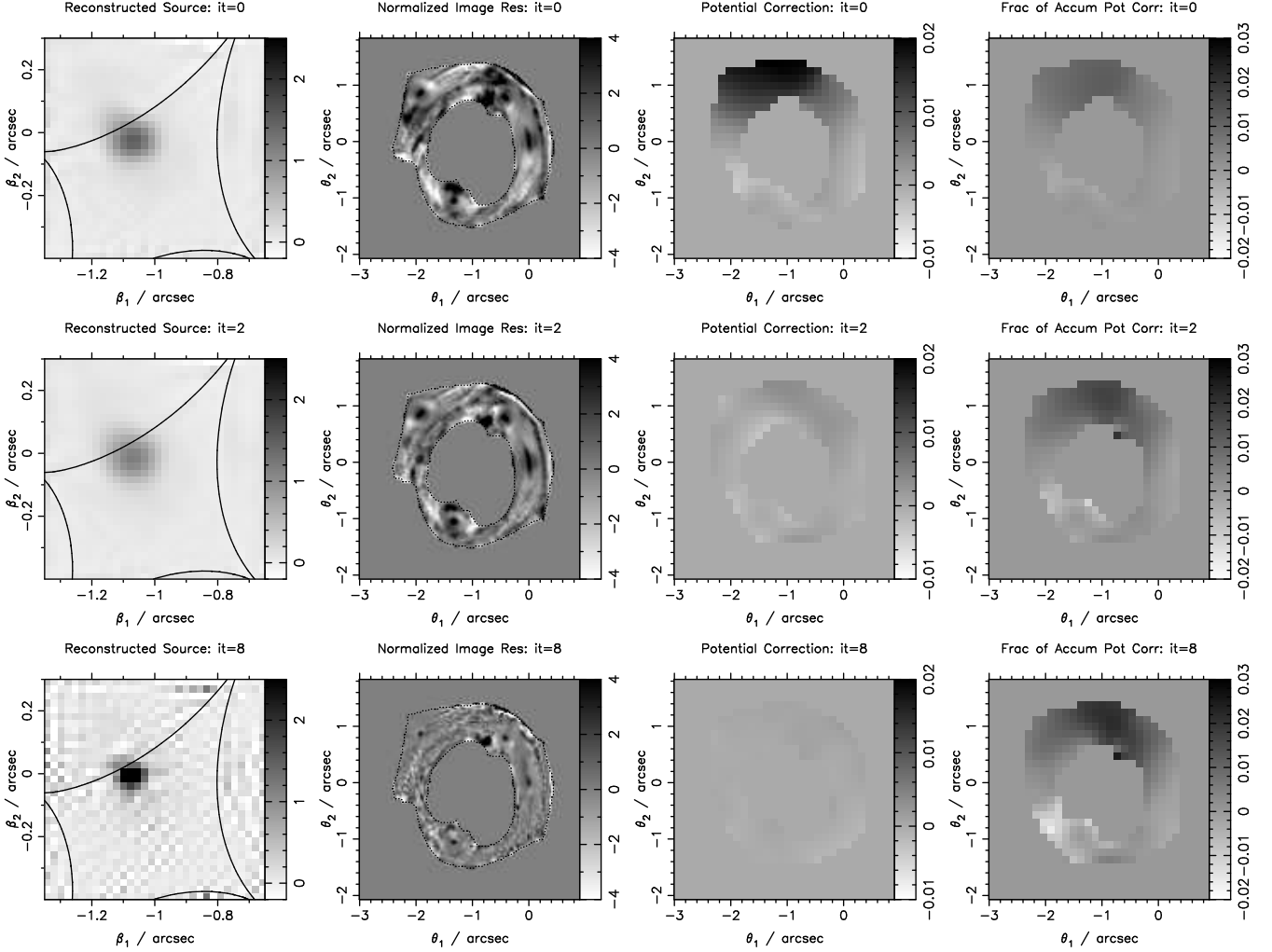

\begin{center}
\includegraphics[width=180mm]{f10a.ps}
\includegraphics[width=180mm]{f10b.ps}
\includegraphics[width=180mm]{f10c.ps}
\end{center}
\caption[Pixelated potential reconstruction of
B1608+656]{\label{fig:PotRec:B1608} Results of the iterative pixelated
  potential reconstruction of B1608+656.  Top row, which shows the
  results of iteration=0: the left-hand panel shows the
  over-regularized curvature source reconstruction, the middle-left
  panel shows the normalized image residual (in units of the estimated
  pixel uncertainty from the data image covariance matrix) based on
  the inverted source, the middle-right panel shows the potential
  corrections on an annulus using the curvature form of regularization,
  and the right-hand panel shows the accumulated potential corrections
  relative to the initial potential model.  The source is localized,
  an indication that we are close to the initial model, but not at the
  true potential model because significant image residuals are
  present. Middle row, which shows the results of iteration=2: the
  panels are arranged in the same way as in the top row.  Compared to
  iteration=0, the image residuals and the potential corrections are
  both smaller.  Bottom row, which shows the results of iteration=8:
  the panels are arranged in the same way as in the top row.  The
  resulting source of the corrected potential is more localized than
  that of the uncorrected potential in
  Fig.~\ref{fig:B1608SrRecB1Star3band}, and the image residual
  corresponds to a reduced $\chi^2$ of 1.1.  The accumulated potential
  correction is only $\sim 2\%$.}
\end{figure*}

The top row of Fig.~\ref{fig:PotRec:B1608} shows the results of
iteration=0 of source and potential reconstruction.  The left-hand
panel shows the reconstructed source that has been over-regularized by
a factor of 1000.  The caustics are those of the initial SPLE1+D
(isotropic) model.  The source is localized and compact, a sign that
the initial SPLE1+D (isotropic) potential we started from is close to
the true model.  The middle-left panel shows significant image
residuals that are to be corrected, especially near the cores of the
images due to the over-regularization of the source-intensity
distribution.  The annular region marks the region of data that we use
for the evidence computation in the final iteration of source
reconstruction.  Using the gradient from the reconstructed source and
the intensity deficit, the middle-right panel shows the potential
reconstruction of iteration=0 and the right-hand panel shows the
fraction of the accumulated potential corrections relative to the
initial model.

The middle row of Fig.~\ref{fig:PotRec:B1608} shows the result of
iteration=2 of source and potential reconstruction.  Compared to
iteration=0 that has the same over-regularization factors, the source
reconstruction is slightly smoother, the image residual has decreased,
and the potential correction is not as large.

In the iterations from 3 to 8, the potential corrections are small;
therefore, the source reconstruction and image residual change only
gradually during these iterations.  The bottom row of
Fig.~\ref{fig:PotRec:B1608} shows the results of iteration=8 (the last
iteration).  The reconstructed source in the left-hand panel has more
background noise than iteration=2 because the source is now optimally
regularized.  The source after the potential correction is more
localized than that before the potential correction in
Fig.~\ref{fig:B1608SrRecB1Star3band}, which is a good indication that
the reconstructed potential is closer to the true potential (up to the
mass-sheet degeneracy).  The normalized image residual in the
middle-left panel shows an overall decrease in the image residual
compared with that in Fig.~\ref{fig:B1608SrRecB1Star3band}.  There
remains intensity deficit near the image locations since the
intensities of point-like images do not generally match due to the
time delays and variability.  This misfit can also be due to the
undersampling of the PSF.  There is also remaining image residual near
image C that is likely due to imperfections in the dust correction.
Nonetheless, the reduced $\chi^2$ inside the annulus is 1.1 
(keeping in mind the unscaled
nature of our image pixel uncertainties).  The right-hand panel in the
bottom row of Fig.~\ref{fig:PotRec:B1608} shows that the final
accumulated potential correction relative to the initial model is
only $\sim 2\%$.  The structure of the accumulated potential
correction may seem to resemble the simulation in
Fig.~\ref{fig:PR:demo1:simData}; however, this does not mean that the
potential correction in B1608+656 corresponds to a mass clump as in
the simulation.  We point out that the maps of the potential
corrections that generally look similar (due to the fixing of the
three points in the annulus) may lead to very different convergence
maps.

The potential reconstruction described above is for Model 5.  After
repeating the procedure for the other models, we find that the image
residual and source reconstruction in the final iteration for the
other three-band dust models are similar in feature to Model 5.  In
contrast, the two-band dust maps' source reconstruction continue to show
nonlocalized source intensities with spurious light pixels outside of
the main component.  Furthermore, parts of the artificial surpluses or
deficits of the dust-corrected light near the color boundaries remain
after the potential correction.  For Model 11 with no dust, the
potential corrections lead to a localized source with image residuals
that show misfit only near the image cores and locations of the dust
lane.

These results of the potential reconstructions can be quantified using
the Bayesian evidence values from the source reconstruction of the
final corrected potential.  Table
\ref{tab:B1608SingleSrRecEvidAfterPPR} lists the evidence values for
Models 2 to 11.  The uncertainties in the evidence values are due to
the source pixelization and the possible range of over-regularization
for the source-intensity reconstruction and lens potential correction.
We explored over-regularization factors in the range between 1 and
1000 for the source intensity, and various factors within 30 of the
regularization constant $\mu$ that corresponds to the peak of $\mu
E_{\dpsi}$ for the potential correction.  The table shows that the
three-band dust maps are consistently ranked higher than the two-band dust
maps, indicating the importance of including the NICMOS image for the
dust correction.  All three-band dust maps give the same evidence values
within the uncertainties, indicating that the various PSF and three-band
dust models are all acceptable.  Furthermore, the resulting Fermat
potential differences between the images for these models agree within
the uncertainties.  Model 11 with no dust leads to the same evidence
value as the values for three-band dust models.  The predicted Fermat
potential differences between the images for Model 11 are also in
similar ranges as those of the three-band dust models.  This shows that
the global structure of the lens potential remains relatively intact
after the dust correction to give similar predicted Fermat potential
values, even though local pixelated potential corrections are flexible
enough to mimic the effects of dust extinction.  It is encouraging
that the dust extinction in B1608+656 does not alter the surface
brightness in a systematic way as to change the global structure of
the lens potential.  This robustness in the global structure of the
lens potential is important for inferring the value of the Hubble
constant.

\begin{table}
\begin{center}
\caption[Ranked model comparison after potential reconstruction]{\label{tab:B1608SingleSrRecEvidAfterPPR} Ranked model comparison after potential reconstruction
}
\begin{tabular}{cccc}
\tableline
\tableline
Model & PSF & dust & log evidence  \\
      &     &      & $(\times10^4)$ \\
\tableline
5 & B1 & Three-band    & $1.77\pm0.05$ \\
9 & C  & B1/three-band & $1.76\pm0.04$ \\
3 & C  & Three-band    & $1.76\pm0.05$ \\
11 & B1 & ---      & $1.76\pm0.05$ \\
2 & drz& Three-band    & $1.75\pm0.05$ \\
7 & B2 & Three-band    & $1.75\pm0.05$ \\
10& B1 & C/two-band  & $1.61\pm0.05$ \\
4 & C  & Two-band    & $1.58\pm0.05$ \\
6 & B1 & Two-band    & $1.41\pm0.05$ \\
8 & B2 & Two-band    & $1.40\pm0.05$ \\
\tableline
\end{tabular}
\end{center}
\tablecomments{In the PSF column, ``drz'' = drizzled TinyTim, ``C'' = closest star, ``B1'' = bright star \#1, and ``B2'' = bright star \#2.  In the dust map column, ``two-band'' represents the dust map obtained from just the two ACS bands, and ``three-band'' represents the dust map obtained from the two ACS and the one NICMOS band.  The uncertainty in the log evidence from the source-intensity reconstruction is due to the source pixelization and the possible range of over-regularization for the source-intensity reconstruction and lens potential correction.  Within the uncertainties, Models 5, 9, 3, 11, 2 and 7 have the highest evidence values.  Note that the three-band dust maps are ranked higher than the two-band dust maps.}
\end{table}

In summary, for the top PSF, dust, and lens galaxies' light models, the
pixelated potential correction scheme was successfully applied to
B1608+656 leading to potential corrections of $\sim 2\%$.  This is
only a small amount of correction, indicating that the smooth
potential model in \citet{KoopmansEtal03} is remarkably good.  The
resulting source is also well localized.

This completes the dissection of the gravitational lens B1608+656.
The image residual is not fully eliminated possibly due to imperfect
PSF, dust, lens galaxies' light modeling, variability in the point
source intensities, finite source resolution, and/or undersampled PSF.
In Paper II, we use the models in Table
\ref{tab:B1608SingleSrRecEvidAfterPPR} to derive $H_0$ and to estimate
its uncertainty associated with the modeling.


\section{Mass and light in the B1608+656 lens system}
\label{sec:B1608prop}

The clean dissection of the lens system in the previous sections
allows us to study the mass and light in G1 and G2.

Since the amount of potential correction is small, we can safely
neglect the implied corrections when estimating the mass associated
with the lens galaxies.  Integrating the SPLE1+D (isotropic) surface
mass density of each of the lens galaxies within their respective
Einstein radii, the mass of G1 enclosed within $r_{\rm{E;G1}}=0.81''$
is $M_{\rm{G1}}=1.9\times10^{11} h^{-1}\, \rm{M_{\sun}}$, and the mass
of G2 enclosed within $r_{\rm{E;G2}}=0.28''$ is
$M_{\rm{G2}}=2.8\times10^{10} h^{-1}\, \rm{M_{\sun}}$ .  Our dust
correction enables us to recover the intrinsic luminosity of the lens
galaxies.  We use the fitted S\'ersic light profiles to estimate the
luminosity of G1 and G2.  Integrating the flux of G1 and G2 within
$r_{\rm{E;G1}}$ and $r_{\rm{E;G2}}$, respectively, the \textit{total
  mass} to rest-frame B-band light ratio of G1 is $({\rm M}/L_{\rm
  B})_{\rm{G1}}=(2.0 \pm 0.2) h\, \rm{M_{\sun}\, L_{\rm B,\sun}^{-1}}$
and of G2 is $({\rm M}/L_{\rm B})_{\rm{G2}}=(1.5\pm 0.2) h\, \rm{M_{\sun}\,
  L_{\rm B,\sun}^{-1}}$.  The total mass and M/L of G1 are
consistent with those from earlier works on B1608+656
\citep[e.g.,][]{FassnachtEtal96} after taking into account the
difference in the Einstein radius (due to the different number of
components in the lens model) and the lowered M/L as a result of the
dust correction.  The M/L ratio of G1 is low compared to the lens
galaxies in \citet{TreuKoopmans04}, which have M/L in the range
$\sim $3--8$\, \rm{M_{\sun}/L_{\rm B,\sun}}$.  This is consistent with
the spectrum of G1 showing signatures of both young and poststarburst
populations, since these types of galaxies can have lower M/L ratios
by a factor of $\sim 10$ compared to other E/S0 galaxies at similar
redshifts \citep[e.g.,][]{vanDokkumStanford03}.  Therefore, even
though B1608+656 consists of two interacting galaxy lenses that lie in
a group \citep{FassnachtEtal06}, the M/L ratio of G1 is consistent
with those in noninteracting lens systems.


\section{Conclusions}
\label{sect:concl}

In this paper, we have described and tested an iterative and
perturbative lens potential reconstruction scheme whose accuracy in
the recovered lens potential is in principle solely limited by the
noise in the data, provided we have extended sources giving
well connected ring-like images.  The method is based on a Bayesian
analysis, which provides a quantitative approach for comparing
different models of the various constituents of a lens system: PSF,
dust, lens galaxy light, and lens potential.  We applied this method
to the gravitational lens B1608+656 with deep \HST ACS observations.
We presented an image processing technique for obtaining a suite of
PSF, dust, and lens galaxies' light models, and compared these models
quantitatively.  For each model, we reconstructed the lens potential
on a grid of pixels, using the simply-parameterized SPLE1+D
(isotropic) model in \citet{KoopmansEtal03} as our initial model.  The
reconstructions for the models with three-band dust maps were deemed
successful in that they led to an acceptable level of image residual
and a well-localized inferred source-intensity distribution.

From our analysis, we draw the following conclusions.

\begin{enumerate}

\item The potential reconstruction method, which simultaneously determines
the extended source intensity and the lens potential distributions on grids
of pixels, can correct for potential perturbations that are $\lesssim5\%$.

\item The mass-sheet degeneracy is broken in the potential corrections by
choosing forms of regularization that suppress large deviations from the
initial (mass-constrained) model unless the data require them.

\item The NICMOS F160W image is needed to complement the ACS F606W and F814W
images for dust correction in order to avoid systematic errors.

\item The level of potential correction required in B1608+656 was found to
be $\sim 2\%$, validating the use of the simply-parameterized model of
\citet{KoopmansEtal03}.

\item The effect of dust extinction does not alter the global structure of
the lens potential, and hence the predicted Fermat potential differences
between the images.

\item The mass and ${\rm M}/L_{\rm B}$ of G1 inside $r_{\rm E}=0.81''$ are
$1.9\times10^{11} h^{-1}\, \rm{M_{\sun}}$ and $(2.0\pm0.2) h \,
\rm{M_{\sun}\, L_{\rm B,\sun}^{-1}}$, respectively.  These values are
consistent with the spectral type of this galaxy, and previous less
accurate estimates of its M/L ratio.

\end{enumerate}

Although the pixelated potential reconstruction method can be applied to any
lens system with an extended source-intensity distribution, it is
particularly useful for measuring $H_0$ in time-delay lenses.  B1608+656 is
the only four-image gravitational lens system that have all three
independent relative time delays measured with errors of a few percent
\citep{FassnachtEtal99, FassnachtEtal02}.  However, current and future
imaging surveys (such as the Canada-France-Hawaii Telescope (CFHT) Legacy Survey,
the Panoramic Survey Telescope \& Rapid Response System, the Large Synoptic
Survey Telescope, and the Joint Dark Energy Mission) either are or soon  will be producing many more lenses:
we can anticipate building up a sample of lens systems that can be 
fruitfully studied using the methods we have developed.


\acknowledgments We thank M. Brada{\v c}, J. Krist, R.
Massey, C. Peng, J. Rhodes, and P. Schneider for useful
discussions and the anonymous referee for helpful comments 
that improved the presentation of the paper.  
We are grateful to M. Brada{\v c} and T.
Schrabback for their help with the image processing.  S.H.S. thanks the
Kavli Institute for Theoretical Physics for the Graduate Fellowship in
the fall of 2006 and for hosting the gravitational lensing workshop,
during which significant progress on this work was made.  S.H.S.
acknowledges the support of the NSERC (Canada) through the
Postgraduate Scholarship.  C.D.F. and J.P.M. acknowledge support under the \HST
program \#GO-10158. Support for program \#GO-10158 was provided by
NASA through a grant from the Space Telescope Science Institute, which
is operated by the Association of Universities for Research in
Astronomy, Inc., under NASA contract NAS 5-26555.  C.D.F. and J.P.M.
acknowledge the support from the European Community's Sixth Framework
Marie Curie Research Training Network Programme, contract no.
MRTN-CT-2004-505183 ``ANGLES.''  L.V.E.K. is supported in part through an
NWO-VIDI career grant (project number 639.042.505).  T.T. acknowledges
support from the NSF through CAREER award NSF-0642621, by the Sloan
Foundation through a Sloan Research Fellowship, and by the Packard
Foundation through a Packard Fellowship.  This work was supported in
part by the NSF under award AST-0444059, the Deutsche Forschungsgemeinschaft under
the project SCHN 342/7--1 (S.H.S.), the TABASGO foundation in the
form of a research fellowship (P.J.M.), and by the US Department of
Energy under contract number DE-AC02-76SF00515.  Based in part on
observations made with the NASA/ESA \textit{Hubble Space Telescope},
obtained at the Space Telescope Science Institute, which is operated
by the Association of Universities for Research in Astronomy, Inc.,
under NASA contract NAS 5-26555. These observations are associated
with program \#GO-10158.

 
\bibliographystyle{apj}
\bibliography{B1608acsAnalysis}

\appendix
\section{The matrix operator for pixelated potential correction}
\label{app:PRmatrix}

A comparison of the potential correction Equation (\ref{eq:pertEq}) with
its matrix form in Equation (\ref{eq:pertEqMat}) shows that the matrix
operator $\PRmatSet$ needs to include the PSF blurring, the
reconstructed source-intensity gradient, and the gradient operator
that acts on the potential perturbations $\dpsiVec$.  We will consider
each of these in the reverse order.

Before discussing the gradient operator, we need to define the domain
over which the gradient operates.  Recall that the potential corrections
are obtained on an annular region that contains the Einstein ring of
the lensed source.  This region was obtained by tracing all the
potential pixels back to the source plane (from the lens equation) and
seeing which ones land on the finite source region of reconstruction.
Only these potential pixels that trace back to the finite source
region will have values of the source-intensity gradient for potential
correction via Equation (\ref{eq:pertEq}).  These pixels tend to mark
an annular region.  Therefore, we need to find the gradient operator on
this annular region for $\dpsi$.

\begin{figure}
\begin{center}
\includegraphics[width=80mm]{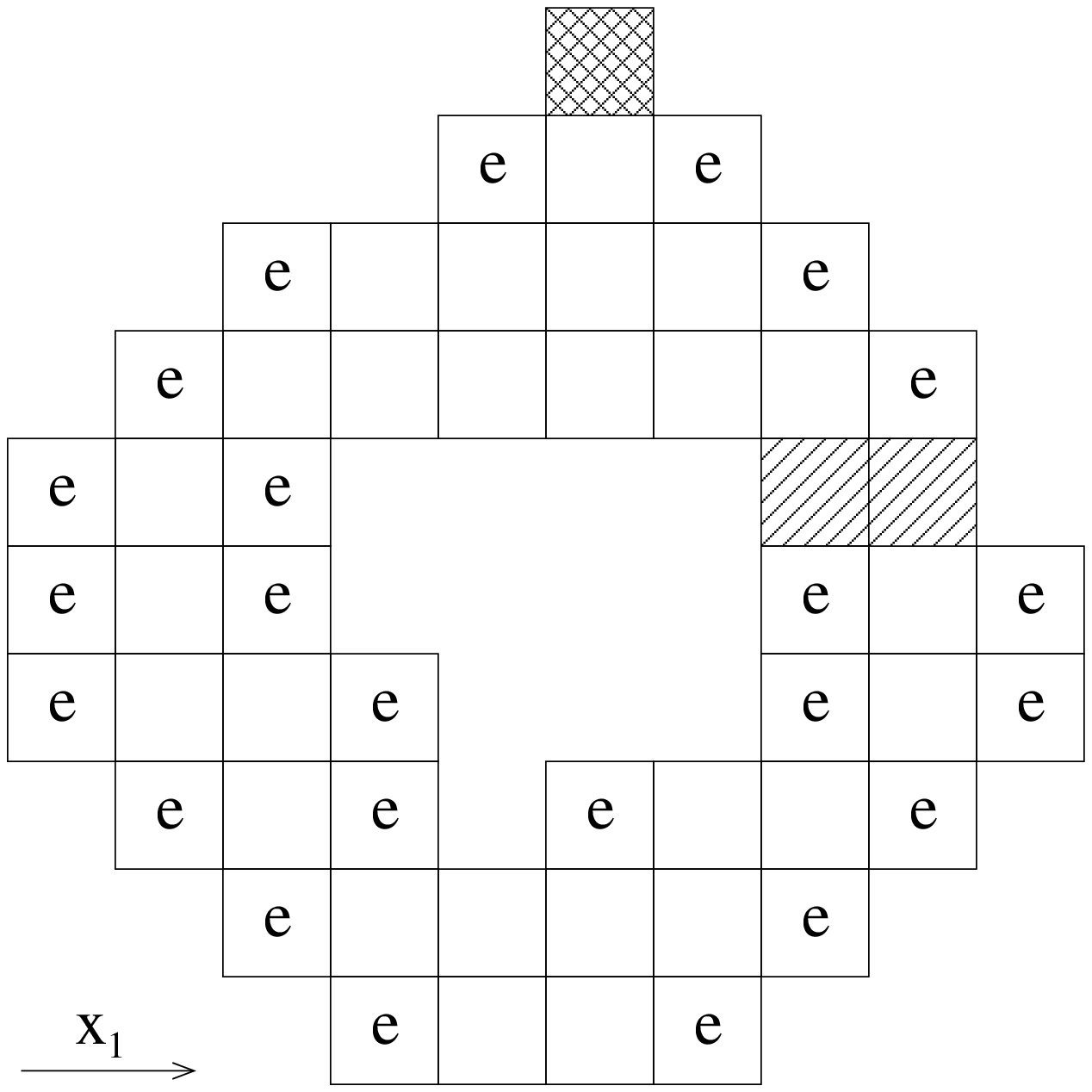}
\end{center}
\caption[Numerical derivative on the annular region of potential
reconstruction]{\label{fig:App:annulus} Typical annular region for
  potential corrections and the form of $\partial f_{i,j}/\partial
  x_1$ for each pixel.  The blank pixels use the $i=2,\ldots,M-1$
  expression for $\partial f_{i,j}/\partial x_1$ in Equation
  (\ref{eq:NumDerRectGrid}).  The pixels with ``e'' are edge pixels that
  use the $i=1$ or $i=M$ expressions for $\partial f_{i,j}/\partial
  x_1$ in Equation (\ref{eq:NumDerRectGrid}).  The shaded pixels use
  Equation (\ref{eq:NumDer2pix}) for $\partial f_{i,j}/\partial x_1$.
  The hashed pixel is an example of an ``exposed'' pixel with no
  adjacent pixel in the $x_1$ direction.}
\end{figure}

To construct the gradient operator, we use finite differencing to
obtain numerical derivatives.  For simplicity, first consider a $M
\times N$ rectangular grid with $x_1$ and $x_2$ as axes and $(i,j)$ as
pixel indices (typically $M\sim N\sim 30$).  In this case, the partial
derivatives of a function $f_{i,j}$ defined on the grid are:
\bea
\label{eq:NumDerRectGrid}
\frac {\partial f_{i,j}}{\partial x_1} & = & \left\{ \begin{array}{ll} \frac{1}{2\Delta x_1}(-3f_{1,j}+4f_{2,j}-f_{3,j}) & \textrm {if $i=1$}\\
\frac{1}{2\Delta x_1}(f_{i+1,j}-f_{i-1,j}) & \textrm {if $i=2,\ldots,M-1$}\\
\frac{1}{2\Delta x_1}(f_{M-2,j}-4f_{M-1,j}+3f_{M,j}) & \textrm {if $i=M$} \end{array} \right. \nonumber \\
\frac{\partial f_{i,j}}{\partial x_2} & = &\left\{ \begin{array}{ll} \frac{1}{2\Delta x_2}(-3f_{i,1}+4f_{i,2}-f_{i,3}) & \textrm {if $j=1$}\\
\frac{1}{2\Delta x_2}(f_{i,j+1}-f_{i,j-1}) & \textrm {if $j=2,\ldots,N-1$}\\
\frac{1}{2\Delta x_2}(f_{i,N-2}-4f_{i,N-1}+3f_{i,N}) & \textrm {if $j=N$} \end{array} \right. ,
\eea
where $\Delta x_1$ and $\Delta x_2$ are, respectively, the pixel sizes in the $x_1$
and $x_2$ directions.  For the annular region of potential
corrections, we only need to elaborate slightly on Equation
(\ref{eq:NumDerRectGrid}).  Fig.~\ref{fig:App:annulus} shows a typical
annular region and the types of pixels when numerically
differentiating in the $x_1$ direction.  The edge pixels of the
annulus, which are denoted by ``e'' in the figure for the $x_1$
direction, are treated as though they are like the edge pixels of the
rectangular grid (so that the $i=1$, $i=M$, $j=1$, or $j=N$ expressions
are used) when the edge pixels are adjacent to at least two other
pixels in the annulus in the direction of which the numerical
derivative is taken.  If an edge pixel of the annulus is only adjacent
to one other pixel in the direction of which the numerical derivative
is taken, such as the shaded pixels in the figure for the $x_1$
direction, then we construct the gradient by taking the difference
between the two and dividing by the pixel size.  For example, if
$f_{i,j}$ is at the edge, and $f_{i+1,j}$ is also in the annulus
(which will have to be an edge pixel if $f_{i+2,j}$ is not in the
annulus), then the numerical derivatives in the $x_1$ direction for
both $f_{i,j}$ and $f_{i+1,j}$ are
\be
\label{eq:NumDer2pix}
\frac{\partial f_{i,j}}{\partial x_1} = \frac{f_{i+1,j}-f_{i,j}}{\Delta x_1}.
\ee
A similar equation applies for the $x_2$ direction.  If an edge pixel
in the annulus is ``exposed'' in the sense that in one of the
directions $x_1$ or $x_2$, it has no adjacent pixels in the annulus,
then this pixel is removed from the annular region of reconstruction
as no numerical derivative can be formed.  An example of an
``exposed'' pixel in the $x_1$ direction is the hashed pixel in the
figure.  Following the above prescription, we can obtain the values
$({\partial f_{i,j}}/{\partial x_1}, {\partial f_{i,j}}/{\partial
  x_2})$ of all the $(i,j)$ pixels in the annulus in terms of values of
the function in the annulus $f_{kl}$.  Factoring out the $f_{kl}$
values, we obtain the gradient operator defined as two matrices:
$\boldsymbol{\mathsf{D}}_1$ for ${\partial }/{\partial x_1}$ and
$\boldsymbol{\mathsf{D}}_2$ for ${\partial }/{\partial x_2}$.

To conform to the data grid (since the image residual and image
covariance matrix is defined on the data grid), we use bilinear
interpolation.  We overlay the data grid on the coarser grid, and for
every data pixel that lies inside the annular region on the coarse
grid, we bilinearly interpolate to get, effectively, gradient
operators on the data grid.  This gives us an $N_{\rm d} \times N_{\rm
  p}$ matrix $\boldsymbol{\mathsf{G}}$ where each row (corresponding
to a data pixel that lies within the annulus) has four nonzero values
that correspond to the coefficients of bilinearly interpolating among
the four coarse potential pixels surrounding this data pixel.
Associated with each data pixel are the source-intensity gradient
values (${\partial I}/{\partial \beta_1}$ and ${\partial I}/{\partial
  \beta_2}$) that were obtained by mapping the data pixel back to the
source plane using the lens equation, and interpolating on the
reconstructed source-intensity gradient on the source grid.  We define
matrices $\boldsymbol{\mathsf{G}}_1$ and $\boldsymbol{\mathsf{G}}_2$
as the matrix $\boldsymbol{\mathsf{G}}$ multiplied by the source-intensity 
gradient components ${\partial I}/{\partial \beta_1}$ and
${\partial I}/{\partial \beta_2}$, respectively.  By definition,
$\boldsymbol{\mathsf{G}}_1$ and $\boldsymbol{\mathsf{G}}_2$ are also
$N_{\rm d} \times N_{\rm p}$ matrices.

Lastly, we represent the PSF as a blurring matrix (operator)
$\blurSet$ that is of dimensions $N_{\rm d} \times N_{\rm d}$ (see
e.g., Section \ref{sec:PPRMethod:realData}; \citet{TreuKoopmans04}).
Note that this matrix $\blurSet$ is different from the matrix in
Section \ref{sec:PPRMethod:matrix:probTheory} that is the Hessian of
the $E_{\rm D}$.

Combining all the pieces together, the matrix operator $\PRmatSet$ is 
\be
\label{eq:PRmatSetExpression}
\PRmatSet = \blurSet \cdot \boldsymbol{\mathsf{G}}_1 \cdot \boldsymbol{\mathsf{D}}_1 + \blurSet \cdot \boldsymbol{\mathsf{G}}_2 \cdot \boldsymbol{\mathsf{D}}_2,
\ee 
which is of dimensions $N_{\rm d}\times N_{\rm p}$.

For the gravitational lens system B1608+656, we also need to include
the effects of dust extinction, which we express as a diagonal matrix
$\dustSet$.  Tracing back along the light rays, we encounter the dust
immediately after the PSF blurring (for the light from the lensed
source).  Therefore, we include it in Equation
(\ref{eq:PRmatSetExpression}) after $\blurSet$ to get the following
expression for the matrix operator $\PRmatSet$ that includes dust:
\be
\label{eq:PRmatSetDustedExpression}
\PRmatSet = \blurSet \cdot \dustSet \cdot \boldsymbol{\mathsf{G}}_1 \cdot \boldsymbol{\mathsf{D}}_1 + \blurSet \cdot \dustSet \cdot \boldsymbol{\mathsf{G}}_2 \cdot \boldsymbol{\mathsf{D}}_2.
\ee


\end{document}